\documentclass[%preprint,
reprint,amsmath,amssymb,aps,jcp,floatfix,]{revtex4-1}

\usepackage{graphicx}
\usepackage{float}
\usepackage{amsmath,amssymb}
\usepackage{xcolor}
\usepackage[utf8]{inputenc}

\usepackage{lineno,hyperref}

\begin{document}

\title{Analytical and computational study of cascade reaction processes in catalytic fibrous membranes}

\author{Gabriel Sitaru}
\author{Stephan Gekle}
\affiliation{Biofluid Simulation and Modeling, Theoretische Physik VI, Universit{\"a}t Bayreuth, Germany}

\begin{abstract}
Multistep catalytic reactions use two different catalysts for the $A\to B$ and the subsequent $B\to C$ reaction, respectively.
Often the employed catalysts are chemically incompatible, such as acid-base systems, which prohibits simple mixing in one solution.
In this work, we study the efficiency of reactors where the incompatible catalytic sites are immobilized on fibrous membranes. 
We compare a lattice Boltzmann based solver for the advection-diffusion-reaction equation, a random walk particle tracking method and a simple theoretical model to investigate the reaction efficiency as a function of two dimensionless control parameters: the P\'eclet and the Damk\"ohler number. 
We find that, while the efficiency decreases with higher flow speed (due to the reduced reaction time), the total production nevertheless increases due to the higher mass flux in most cases.
Our results further show that, even at high flow speeds, spatial proximity of the two catalysts increases reaction efficiency, which supports recent experimental efforts to locate both catalysts on a single fiber in a side-by-side geometry.
\end{abstract}

\maketitle

\section{Introduction}
\label{sec:intro}
Cascade chemical reactions play an important role in nature and many technological applications \cite{Lu_2015_catalysis, Wheeldon_2016, Zeng_Qiang_2016}.
Here, an initial species $A$ reacts to the final product $C$ via an intermediate species $B$ where the two involved reactions $A\to B$ and $B \to C$ each require a different catalyst. 
Unfortunately, in many cases the two catalysts are chemically incompatible (e.g. acid and base) thus preventing their mixing within one pot \cite{Gelman_2000}.
Examples for such reactions include the Knoevenagel \cite{Boucard_2001, Zhang_2016_catalysis} or the Baylis-Hillman \cite{Helms_2005} reaction.
Recently, experiments with electrospun fibrous membranes have demonstrated their great potential as efficient reactor systems for cascade reactions with incompatible catalysts in one-pot \cite{Agarwal_2010_fibers, Pretscher_2019_precise, Pretscher1}.
In this approach, two types of elongated fibers, one containing the first and the other the second catalyst, are combined into a single membrane through which the reactants are flown. 
This setup provides a highly efficient reactor geometry while at the same time preventing direct contact - and thus annihilation - of the two incompatible catalysts.

From the theoretical perspective, these systems constitute an advection-diffusion-reaction system. 
Most work in this area considers the movement by diffusion only \cite{samson1, Castellana_2014_catalysis, Roberts_2015, Li_2016_diffusionReaction}.
Studies which take into account external flow include Brownian dynamics simulations for spherical catalyst geometries \cite{Bauler_2010, Biello_2015}, 2D simulations with reacting boundaries \cite{Arcidiacono_2008, ABDOLLAHZADEH} or microchannels with obstacles \cite{Succi_2001}.

In this work, we focus on predicting the reaction efficiency of one- and two-step reactions when the catalytic sites have a cylindrical geometry and an additional external flow acts to transport the reactants between the various sites.
Using two different simulation methods (Lattice-Boltzmann and a random walk particle model) together with an approximative analytical theory, we predict the efficiency as function of the two relevant dimensionless parameters: the P\'eclet and the Damk\"ohler number.
The former captures the ratio between advective and diffusive transport while the latter measures the ratio between reactive and diffusive time scales.
The theory and the numerical results show a good agreement for a wide range of these dimensionless numbers.

\section{System setup}
\label{sec:system_setup}
In our simulations, the fibers are modeled as infinitely long cylinders arranged in various geometries ranging from a single reaction performed by one fiber to multi-step reactions between two membranes containing randomly arranged fibers.
Similar to a flow-through reactor, a pressure gradient is set from the entrance to the exit of the system in order to generate the velocity field for the fluid.

\subsection{System parameters and dimensionless numbers}
\label{sec:parameters}
The behavior of our systems is characterized by the interplay of three different phenomena - advection, diffusion and reaction - which can be reduced to two dimensionless numbers.
The first is the P\'eclet number 
\begin{align}
\mathrm{Pe} =\frac{u_{0}R}{D}
\end{align}
giving the ratio between the advection velocity $u_0$, the fiber radius $R$ and the diffusion coefficient $D$.
For $u_0$ we use the centerline, i.e., maximum velocity at the entrance.
The second is the Damk\"ohler number
\begin{align}
\mathrm{Da} =\frac{kR^{2}}{D}
\end{align}
giving the ratio between the reactive and the diffusive time scales.
In section~\ref{sec:geo_single} we consider an $A\to B$ reaction followed by a cascade $A\to B \to C$ reaction in \ref{sec:geo_cascade}.
As an output parameter of our study we consider the reaction efficiency 
\begin{alignat}{2}
\epsilon &= \frac{N_{B}}{N_{A}+N_{B}} &&\;\;\;\;\;\; \mathrm{for}\;A\to B\\
\epsilon &= \frac{N_{C}}{N_{A}+N_{B} + N_{C}} &&\;\;\;\;\;\; \mathrm{for}\;A\to B\to C
\label{A-B}
\end{alignat}
where $N_x$ is the number of particles of species $x$ that flow out of the system per time unit. 
As both the system parameters as well as the output quantities are dimensionless and thus independent of the employed unit system, we will use simulation units in the following for simplicity.

\subsection{System geometry: Single reaction}
\label{sec:geo_single}
\begin{figure} [htb]
	%single column image
	\includegraphics[clip, width=0.8\linewidth]{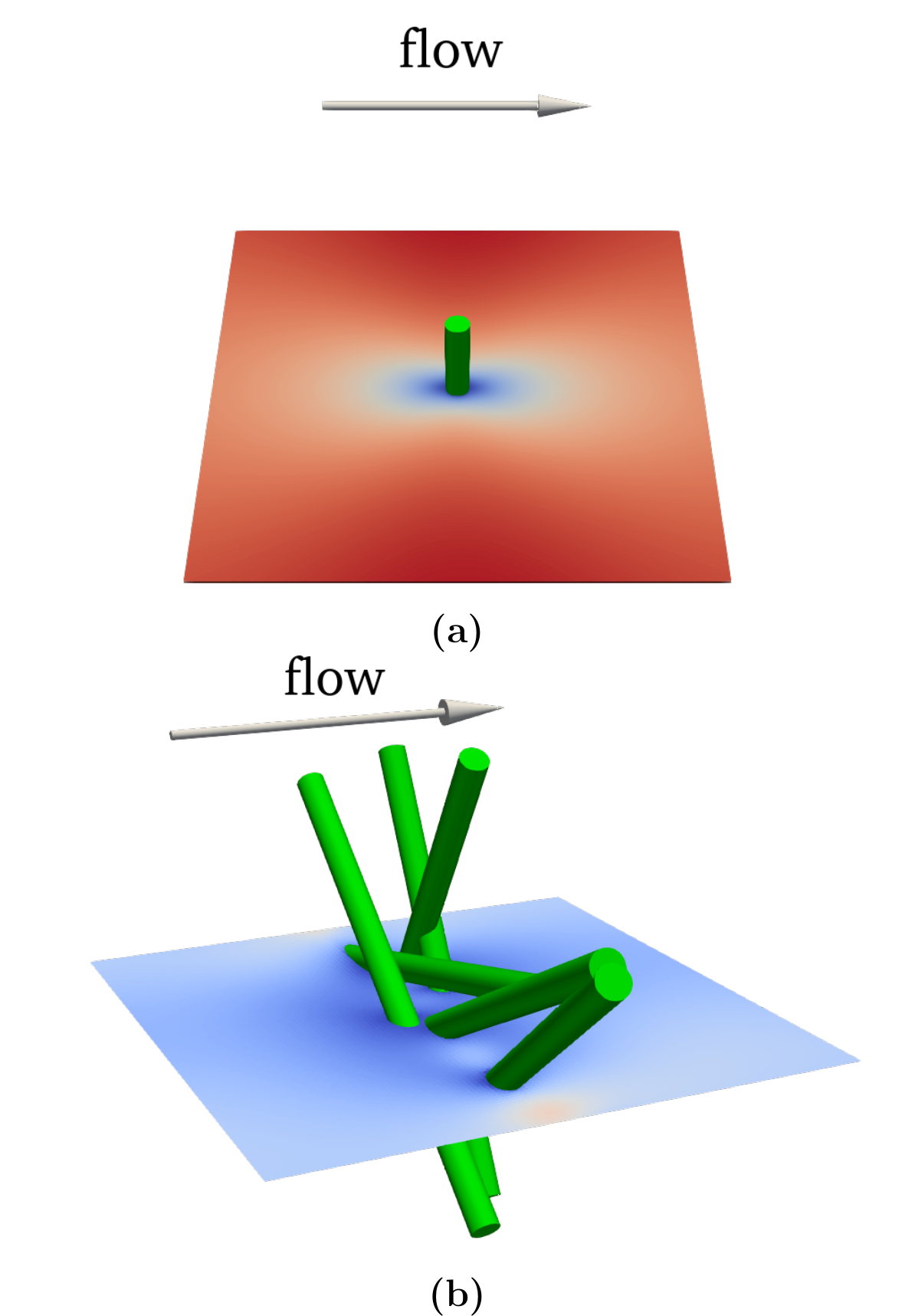}
	\centering
	\caption{Illustration of the system geometries and the velocity field used in our single-reaction studies: 
		(a) an isolated fiber, 	
		(b) a membrane with randomly oriented fibers}
	\label{system_AB}
\end{figure}

We start with an $A\rightarrow B$ reaction performed by a single catalyst.
The catalytic site here has the shape of an isolated cylinder with radius $R=5$ and is placed in the middle of a $300\times300$ box, the third dimension being irrelevant due to the translational symmetry along the fiber axis as illustrated in Fig.~\ref{system_AB}(a).
The next step is to add multiple fibers to replicate a regular fibrous membrane. 
For this configuration, a $120 \times 100 \times 120$ box was used and 6 cylinders having a radius $R=3$ were placed with a random orientation and position as illustrated in Fig.~\ref{system_AB}(b).

% Cascade
\subsection{System geometry: Cascade reaction}
\label{sec:geo_cascade}
\begin{figure*}[htb]
	%2-column image
	\includegraphics[width=0.75\textwidth]{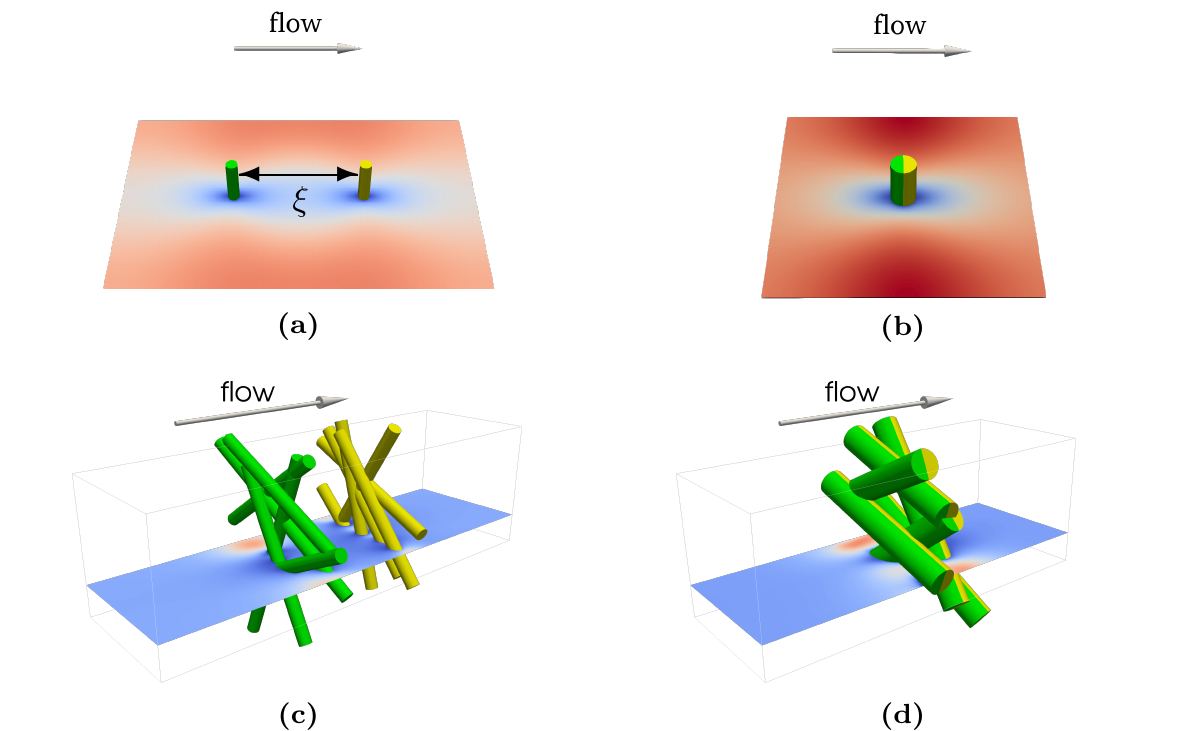}
	\centering
	\caption{Illustration and velocity field for: (a) two fibers, (b) a single side-by-side fiber, (c) two membranes, each carrying a different catalyst, (d) a single side-by-side membrane}
	\label{system_ABC}
\end{figure*}

The central goal of our work is to understand and to predict reaction efficiencies for cascade reactions.
For this, we start with the simplest case for a two-step cascade reaction where two individual fibers act as catalytic sites for the $A\rightarrow B$ and the $B\rightarrow C$ reaction, respectively. 
Fig.~\ref{system_ABC}(a) shows such a system for two fibers having a radius $R=5$ and separated by a distance $\xi=100$ in a box of $1000\times 300$. 

We furthermore introduce an interesting special case termed the side-by-side morphology: here two incompatible catalysts are immobilized next to each other on the same fiber with a common interface running all through the length of the fiber as illustrated in Fig.~\ref{system_ABC}(b) \cite{Chen_2009_nanosprings}.
In order to be able to compare the different systems, we choose the radii such that the total surface area of the catalytic sites remains the same leading to $R=10$ for the single side-by-side fiber.
The box size is $1000\times 300$. 

For both geometries, we then also study the randomly oriented fiber membranes illustrated in Fig.~\ref{system_ABC}(c) and~(d) where the radii are again reduced to $R=3$ (box size $200\times60\times60$) and $R=6$ (box size $160\times60\times60$), respectively.

%
% Simulation methods
%

\section{Simulation methods}
\label{sec:sim_methods}
\subsection{Lattice Boltzmann}
\label{sec:LBM}
The evolution of the time-dependent concentration profile $C_{j}$ of species $j$ throughout the reactor is governed by the advection-diffusion-reaction equations (ADRE)
\begin{equation}
\frac{\partial C_{j}}{\partial t} +\vec{u} \cdot \nabla C_{j}-D_{j}\Delta C_{j}=R_{j} 
\label{ADReq}
\end{equation}
where $\vec{u}$ is the local advection velocity, $D_{j}$ the diffusion coefficient and $R_{j}$ is a reaction term. 
For the $A \rightarrow B$ reaction, the latter assumes the first-order reaction form  
\begin{equation}
\begin{split}
R_\mathrm{A}&=-kC_\mathrm{A}\\
R_\mathrm{B}&=+kC_\mathrm{A}
\end{split}
\label{reaction_term}
\end{equation}
with the reaction rate $k$ and an analogous form for the $B\to C$ reaction.
Since there is no back-coupling of the species concentration to the fluid properties and the flow field, the velocity $\vec{u}$ in equation~(\ref{ADReq}) is constant in time, but not in space.
We therefore employ a hybrid scheme where we first use the Lattice-Boltzmann method to compute the stationary velocity field $\vec{u}$ for a given geometry (see below).
Subsequently, this velocity field is used as input for a second Lattice-Boltzmann method which solves the ADRE equation~(\ref{ADReq}) as described further below.

\subsubsection{Lattice-Boltzmann for the Navier-Stokes equation}
\label{sec:LBM_NS}
Inspired by the original lattice gas model \cite{lattice_gas}, the Lattice Boltzmann Method (LBM) tracks the distribution of particles over a discretized space and time domain with a resolution $\Delta x$ and $\Delta t$, respectively \cite{Krueger}.
The probability distribution function $f(\vec{x},t)$ which, for each lattice node, is discretized into populations according to the number of discrete velocities $\vec{c}_{i}$ and their weight $w_{i}$, obeys the Lattice Boltzmann equation (LBE)
\begin{equation}
f_{i}(\vec{x}+\vec{c}_{i}\Delta t,t+\Delta t)=f_{i}(\vec{x},t)+\Omega_{i}(\vec{x},t)
\label{LBE}
\end{equation}
where $\Omega_{i}$ is the collision operator.
Here we use the LBM implementation in the free software package ESPResSo \cite{espresso, Weik_2019, Bacher_2018} which is based on the D3Q19 grid model and the multiple-relaxation-time collision operator.
The pressure gradient is implemented as a body force $f=10^{-6}$ in $x$ direction.
With a time step of $ \Delta t=1$, the simulation is typically run for 6000 time steps until a steady velocity field is reached.
For simplicity, we simulate each system geometry once with a low pressure gradient and obtain higher velocities by simply multiplying this basic flow field with a constant scaling factor.
Due to the linearity of Stokes flow, this procedure is exact for Stokes flow at $Re\ll 1$, a condition which is satisfied in most of our setups.
At high velocities, i.e. high $Pe$ numbers, the maximum Reynolds number occurring in our systems is $Re\approx 7$, where inertial corrections are expected to be small and linearity is still a reasonable approximation.
The boundaries of the system are periodic in all directions.
At the membrane surfaces a bounce-back boundary condition ensures the no-slip condition.

\subsubsection{Lattice-Boltzmann for Advection-Diffusion-Reaction}
\label{sec:LBM_ADE}
To solve the ADRE, we developed a separate LBM solver in which a source term $Q_{i}(\vec{x},t)$ was added into equation~(\ref{LBE}) to model the chemical reactions \cite{Kang}
\begin{equation}
g_{i}(\vec{x}+\vec{c}_{i}\Delta t,t+\Delta t)-g_{i}(\vec{x},t)=\Omega_{i}(\vec{x},t)+Q_{i}(\vec{x},t)
\label{LBE_ADE}
\end{equation}
As collision operator, we here use the BGK model \cite{BGK}
\begin{equation}
\Omega_{i}(\vec{x},t)=-\frac{1}{\tau_\mathrm{g}}\left( g_{i}(\vec{x},t)-g_{i}^\mathrm{eq}(\vec{x},t)\right) 
\label{BGK}
\end{equation}
where $\tau_{g}$ is the relaxation time and $g_{i}^{eq}$ the equilibrium distribution. 
For the latter, it has been shown \cite{geq} that a good stability is obtained using the Taylor expansion of the Maxwell-Boltzmann equilibrium distribution function up to second order
\begin{equation}
g_{i}^\mathrm{eq}=w_{i}C\left(1+\frac{\vec{c}_{i}\cdot \vec{u}}{c_{s}^2}+\frac{\left( \vec{c}_{i}\cdot \vec{u}\right) ^2}{2c_\mathrm{s}^4}-\frac{\vec{u}\cdot \vec{u}}{2c_\mathrm{s}^2} \right) 
\end{equation}
where the speed of sound for the D3Q19 lattice takes the form $c_s^{2}=\frac{1}{3}\frac{\Delta x^2}{\Delta t^2}$, while the velocity $\vec{u}$ is externally imposed as described above.
The diffusion coefficient $D$ is given by the relaxation time according to
\begin{equation}
D=c_\mathrm{s}^{2}\left(\tau_\mathrm{g}-\frac{\Delta t}{2} \right) 
\label{Diffusion_LBM}
\end{equation}
while the concentration $C$ is defined in terms of the distribution function by
\begin{equation}
C=\sum_{i}g_{i}
\end{equation}
To relate the LBM reaction term with the physical reaction rate $k$, we start from the simple scheme first presented by \citet{Dawson} where the source term is discretized over the lattice nodes \cite{Dawson, Kang, BLAAK2000}
\begin{equation}
\begin{split}
Q_{i}^\mathrm{A}&=-k\Delta t w_{i} C_\mathrm{A}\\
Q_{i}^\mathrm{B}&=+k\Delta t w_{i} C_\mathrm{A}
\end{split}
\label{source_lbm}
\end{equation}
and analogously for the $B \to C$ reaction.
As our aim is to model a surface-catalytic reaction, we use (\ref{source_lbm}) only for those populations that stream into a reactive boundary node. 
For all others, we set $Q_{i}^{A,B}=0$.
A comparison of this approach to the standard situation where all populations are allowed to react is presented in Fig.~\ref{node_vs_pop}.

For the ADE LBM we use a time step of $\Delta t=1$ and a steady situation was obtained after a number of time steps ranging from 130000 in the low $Pe$ regime to 8000 for higher $Pe$. 
The employed lattice was identical to the one used in the NS LBM above.

\subsubsection{Boundary conditions and flux computation in ADE}
\label{sec:LBM_BC}
In ADE, the concentration along a boundary can vary generating a tangential flux while the normal flux must be zero due to impenetrability of the boundary.
This behavior can be recovered from the anti-bounce-back scheme \cite{antiBB} which, for a node $\vec{x}_{b}$ next to a stationary wall, reads
\begin{equation}
g_{\bar{i}}\left( \vec{x}_\mathrm{b}, t+\Delta t\right)=-g_{i}^{\ast}\left( \vec{x}_\mathrm{b}, t\right)+2w_{i}C_{w}
\label{BC}
\end{equation}
where $g_{\bar{i}}$ is the population streaming away from the boundary, $g_{i}^{\ast}$ the post-collision population streaming towards the wall and $C_{w}$ is the imposed wall concentration.

For our ADE simulations, a homogeneous concentration of the initial substance $C_{A}=1$ is imposed at the entrance while for the rest of the species a simple bounce-back is performed.
At the channel outlet the normal derivatives are set to zero for every species, which can be transformed into a Dirichlet condition by setting $C_w=C_b$ in (\ref{BC}), where $C_b$ is the concentration of the node next to the boundary (see 8.5.3.2 of \citet{Krueger}). 

To determine the outflux of a species, we consider a plane perpendicular to the flow direction and located two lattice points away from the right system boundary. 
In this plane, the difference between all the populations streaming towards the outlet (right) and those streaming to the left is computed for each node and subsequently averaged across all nodes.
The concentrations are then directly proportional to the particle numbers $N$ in equation~(\ref{A-B}).
The remainder of the boundaries are considered periodic.

\subsubsection{Membrane generation}
\label{sec:LBM_memb}
Membranes were generated by placing randomly oriented cylindrical fibers. 
Cases where two or more fibers showed significant overlap were discarded.
The cylinders are then included as boundaries as described in Section \ref{sec:system_setup} into the ESPResSo software.
Due to the random orientation, the periodicity of the fibers cannot be ensured.
Therefore, the cylinders are closed at both ends to avoid the fluid running inside.

\subsection{Random walk}
\label{sec:RW}
In addition to the LBM, a random walk particle tracking (RWPT) approach, which is known for its equivalence with the ADRE \cite{randomwalk_perez}, was also used in our study. 
For simplicity, we restrict our application of the RWPT approach to the geometries with translational invariance along the fiber axes illustrated in Fig.~\ref{system_AB}(a) as well as \ref{system_ABC}(a) and~(b).
In our 2D implementation, the movement of each particle is composed of two different contributions due to advection and diffusion.
For the advective contribution, the externally imposed velocity at a particle position is obtained by applying a bilinear interpolation to the discrete flow field generated with a LBM-based solver described in Section~\ref{sec:LBM_NS}.
The particle position is then updated using this interpolated velocity by simple Euler integration with $\Delta t=0.5$.
For the diffusive contribution, the length of the diffusion step is constant in time and computed from the mean squared displacement as $\mid\vec{x}\left(t+\Delta t \right) -\vec{x}\left(t\right)\mid=\sqrt{4D\Delta t}$. 
The direction is randomly chosen.

The collision between a particle and a fiber is considered elastic, i.e. upon collision, the normal component of the particle velocity is inverted while the tangential component is not affected.
To model the reaction, we consider a thin reactive zone of size $\delta=0.2$ around each fiber.
For each particle within this zone, the probability to react within a time step is then $k\Delta t$. 

Our algorithm ensures that a constant homogeneous concentration along the entrance of the channel is set at any time by randomly adding or removing particles where necessary.
The efficiency is computed using the definition given in equation~(\ref{A-B}) by counting the number of particles that leave the reactor for each species during one time step. 
The results of the RWPT model agree very well with the LBM simulations for all investigated situations.

%
% Results
%
\section{Results and theoretical model}
\label{sec:results}
\subsection{Single reaction}
\label{sec:res_single}
\begin{figure} [htb]
	%single column image
	\includegraphics[width=0.8\linewidth]{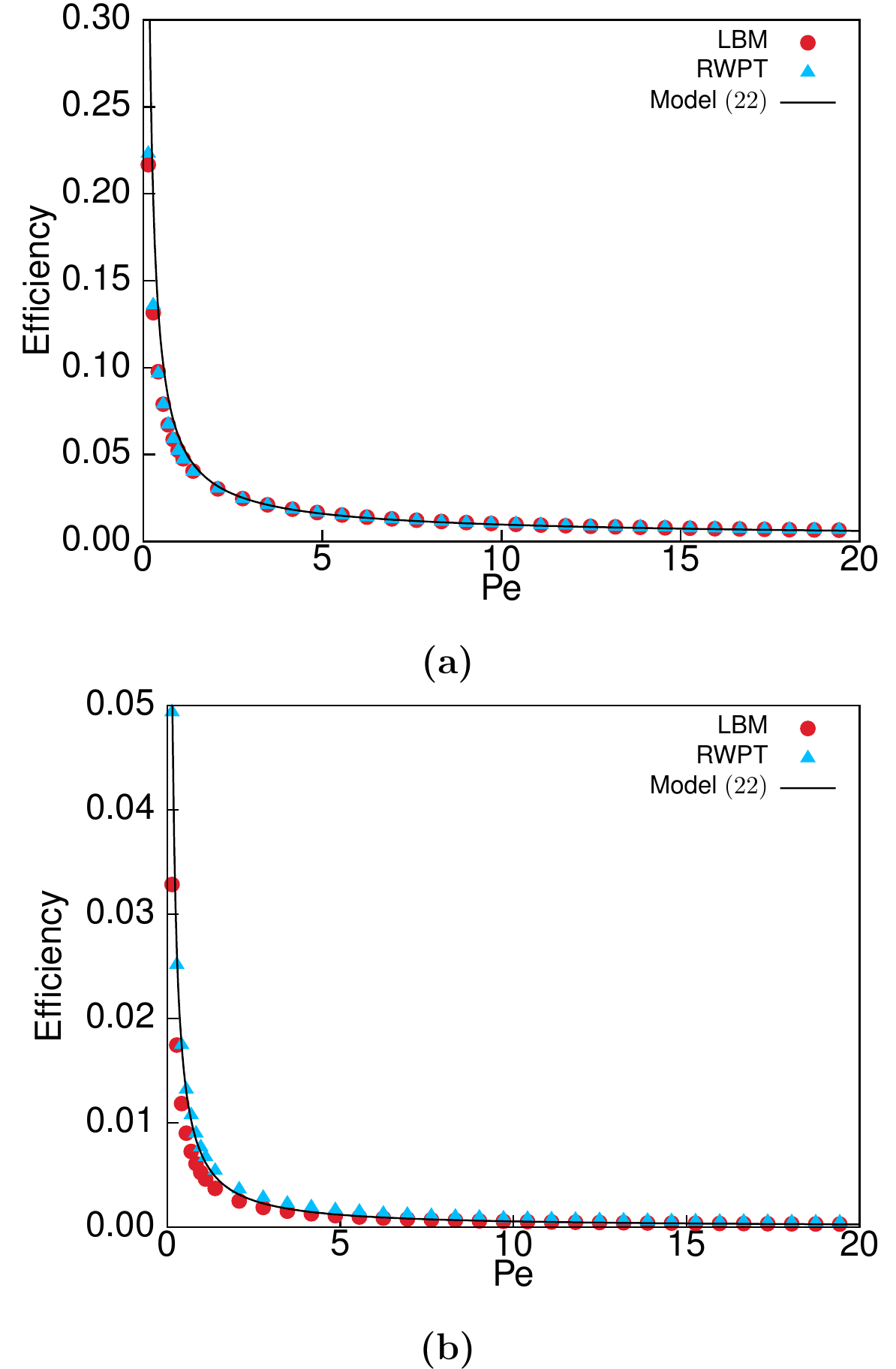}
	\centering
	\caption{Efficiency of an $A\rightarrow B$ reaction as a function of $\mathrm{Pe}$ for the isolated fiber illustrated in Fig.~\ref{system_AB}(a).
		Parameters are (a) $\mathrm{Da=150}$ ($k=1$) and (b) $\mathrm{Da=1.5}$ ($k=0.01$)}
	\label{singleAB}
\end{figure}

We start our study by the setup illustrated in Fig.~\ref{system_AB}(a): a single catalytic fiber is located at the center of a domain where the initial reactant $A$ is advected by a fluid from left to right and, after suffering a reaction on the surface of the fiber, is converted into the product species $B$. 
In Fig.~\ref{singleAB} we show the efficiency defined by equation~(\ref{A-B}) computed from the LBM simulations as well as the RWPT model as function of the P\'eclet number $\mathrm{Pe}$. 
The decrease of the efficiency $\epsilon$ with $\mathrm{Pe}$ can be easily understood: the higher $\mathrm{Pe}$, the higher the advection velocity, i.e. the shorter the time that the reactant spends near the catalyst which lowers the percentage of reacted particles.
As can be seen by comparing Fig.~\ref{singleAB}(a) and~(b) this behavior is qualitatively independent of the reaction rate $k$ (or, equivalently the dimensionless Damk\"ohler number $\mathrm{Da}$).
Fig.~\ref{prod_AB_single}, however, shows that nevertheless the total production increases at higher flow speeds due to the higher mass flux.

\begin{figure} [htb]
	%single column image
	\includegraphics[clip, width=0.8\linewidth]{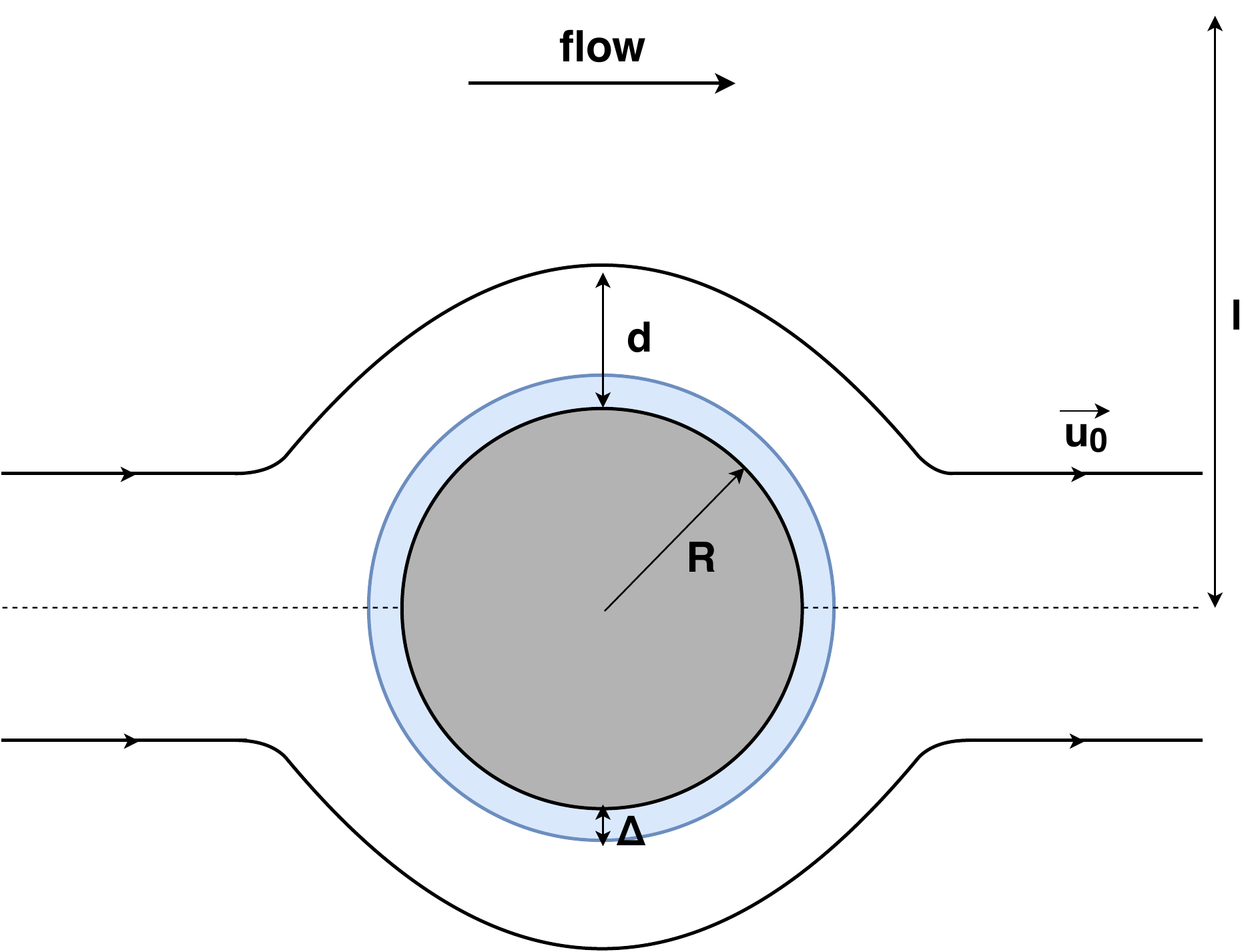}
	\centering
	\caption{Illustration of the approximative theoretical model. The streamline separates the particles of reactant that never reach the cylinder from those that suffer at least one collision with it. The reaction can take place only in the highlighted region at a maximum distance $\Delta$ from the surface of the fiber}
	\label{flow_model}
\end{figure}

We proceed to explain these simulation results by introducing an approximative theoretical model.
In our model, we consider the advection of a substance from left to right through a 2D region with width $l$ around an infinitely long cylinder with radius $R$ as shown in Fig.~\ref{flow_model}. 

We now assume the existence of a region with width $d$ around the cylinder within which molecules of species $A$ are able to collide with the cylinder surface by diffusion.
In contrast, all molecules outside $d$ are advected through the reactor without ever reaching the catalytic surface.
If the cylinder is small compared to the entire system size, we can therefore express the ratio of collided to the total number of molecules simply as the ratio of the widths of the two regions 
\begin{equation}
\epsilon_\mathrm{coll} = \frac{d}{l}
\label{eps_coll1}
\end{equation}
where we have assumed a homogeneous species distribution at the system entrance.
In order to determine the distance $d$, we consider a molecule moving with velocity $\overrightarrow{u_{0}}$ along the streamline that separates the two regions. 
For such a particle to collide, the time $t_{a}$ required to advect past the cylinder is equal to the time $t_{D}$ required to diffuse towards the cylinder:
\begin{equation}
t_\mathrm{a} = \frac{\pi\left(R+d\right)}{u_{0}}
\label{ta}
\end{equation}
\begin{equation}
t_\mathrm{D} = \frac{d^{2}}{2D}.
\label{tD}
\end{equation}
Equating (\ref{ta}) and (\ref{tD}) yields the following quadratic equation for $d$
\begin{equation}
\frac{1}{2D}d^{2}-\frac{\pi}{u_{0}}d-\frac{\pi R}{u_{0}}=0.
\label{d}
\end{equation}
Considering that only half of the particles diffuse towards the cylinder, replacing the solution of (\ref{d}) in (\ref{eps_coll1}) and introducing the P\'eclet number gives using the positive root of (\ref{d}):
\begin{equation}
\epsilon_\mathrm{coll}=\frac{1}{2}\left( \frac{\pi}{\mathrm{Pe}}+\sqrt{\frac{\pi^{2}}{\mathrm{Pe}^{2}}+\frac{\pi}{2\mathrm{Pe}}}\right)\frac{R}{l} ,
\label{eps_coll}
\end{equation}

\begin{figure*} [ht]
	%2-column image
	\includegraphics[width=0.75\textwidth]{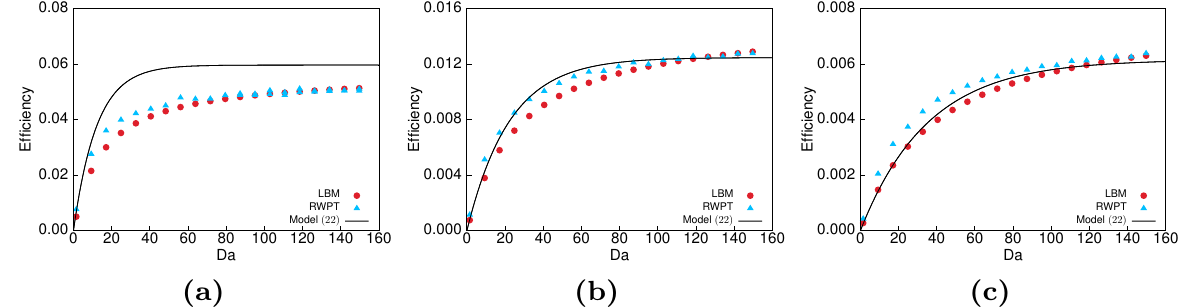}
	\centering
	\caption{Efficiency of an $A\rightarrow B$ reaction with an isolated fiber as a function of Da for (a) $\mathrm{Pe}=1$, (b) $\mathrm{Pe}=7$, (c) $\mathrm{Pe}=20$}
	\label{singleAB_Da}
\end{figure*}

All molecules that reach the cylinder surface react with a reaction rate $k$, thus the efficiency~(\ref{A-B}) can be rewritten as
\begin{equation}
\epsilon=\epsilon_\mathrm{coll}\left( 1-e^{- kt_\mathrm{r}}\right) 
\label{eps2}
\end{equation}
where $t_r$ is the reaction time scale, i.e.\ the time that the molecule spends in close proximity to the catalytic surface.
To estimate $t_r$ we introduce a thin reaction shell of width $\Delta$, see Fig.~\ref{flow_model}, within which the reaction takes place.
The ratio between the reactive and the diffusive time scale can be expressed as the ratio $ \alpha $ between the areas within which the reaction takes place versus the area where collisions take place
\begin{align}
\alpha =\frac{ t_\mathrm{r} } { t_\mathrm{D} }  & =\frac{ \left( R+ \Delta \right)^2 - R^2 } { \left( R+d \right)^2 - R^2 }   \nonumber\\
& \stackrel{ (R\gg d, \Delta) } {  \approx  } \frac{ \Delta  } { d } 
\end{align}
This leads to the efficiency
\begin{equation}
\epsilon=\epsilon_\mathrm{coll}\left( 1-e^{-k\alpha t_ \mathrm{D}}\right).
\label{eps}
\end{equation}
The width of the reaction shell is considered as a fitting parameter and is here chosen as $\Delta =0.5$.
This value is kept constant for all simulation and is clearly sensible: keeping in mind that the LBM algorithm works on a rectangular grid with unit spacing, the average distance between the surface and a neighboring lattice point will be of the order of half a grid cell. 

The model predictions are in very good agreement with the simulation results as can be seen in Fig.~\ref{singleAB}.
Only in the low $\mathrm{Pe}$ regime, certain deviations occur.
These are to be expected as for low $\mathrm{Pe}$ the collision zone $d$ becomes large compared to the system width $l$ and the idealized clear-cut separation between the collision and the no-collision zone on which our model is based becomes increasingly blurred.

We proceed to analyze the dependence of the efficiency on the reaction rate.
The simulation data in Fig.~\ref{singleAB_Da} shows the expected trend that $\epsilon$ increases with $\mathrm{Da}$.
The growth rate slows down with increasing $\mathrm{Da}$ and eventually a plateau is reached where the reaction rate is so high that every molecule that collides with the surface will immediately react and thus no further increase in efficiency will occur.
This behavior is qualitatively and, within some limits, also quantitatively reproduced by the theoretical model.
We note again that the only adjustable parameter is $ \Delta $ which is fixed to $ \Delta =0.5$ throughout the entire manuscript and not re-fitted for each simulation series individually.
The corresponding total production is shown in Fig.~\ref{prod_AB_single_Da}.
%\begin{figure*} [htb]
%	%2-column image
%	\includegraphics[width=0.75\textwidth]{Fig5}
%	\centering
%	\caption{Efficiency of an $A\rightarrow B$ reaction with an isolated fiber as a function of Da for (a) $\mathrm{Pe}=1$, (b) $\mathrm{Pe}=7$, (c) $\mathrm{Pe}=20$}
%	\label{singleAB_Da}
%\end{figure*}

\begin{figure} [htb]
	%single column image
	\includegraphics[width=0.8\linewidth]{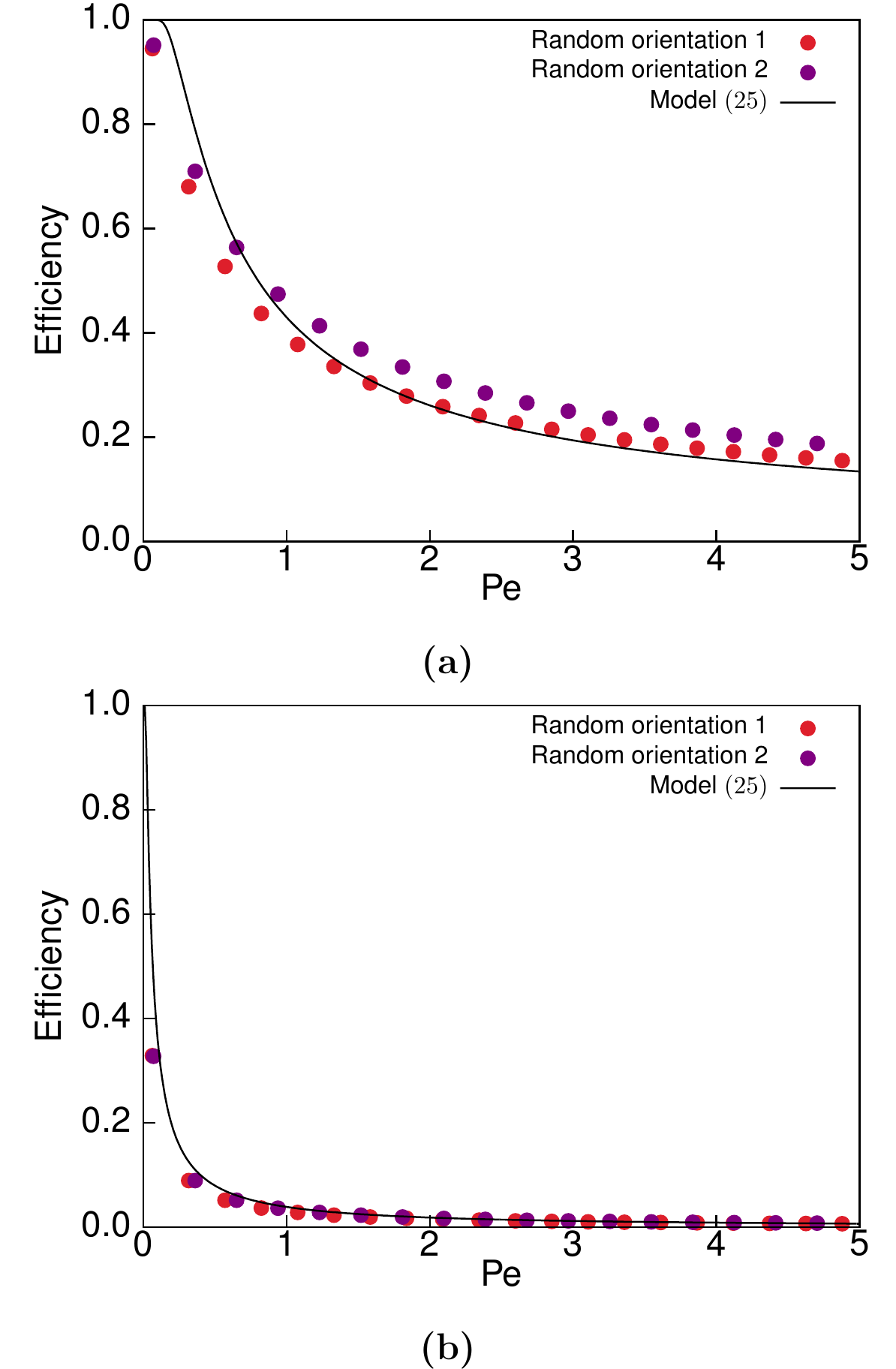}
	\centering
	\caption{Efficiency of a single reaction performed by a  membrane consisting of six randomly oriented fibers as illustrated in Fig.~\ref{system_AB}(b).
		(a) $\mathrm{Da}=150$, (b) $\mathrm{Da}=1.5$}
	\label{figABmemb}
\end{figure}

Having understood the system behavior for an isolated fiber, we proceed to study the behavior of multiple fibers.
If the fibers are close to each other, this is a non-trivial extension as catalytic centers can scavenge each other's reactants thus reducing the overall efficiency of the system.
We consider two membranes each consisting of six randomly aligned fibers as illustrated in Fig.~\ref{system_AB}(b).
From the Lattice-Boltzmann data shown in Fig.~\ref{figABmemb} we find that the general trend (decrease of $ \epsilon$ while increasing $\mathrm{Pe}$) is the same as for the isolated fiber, but that the drop in efficiency appears to be less drastic than in the isolated fiber scenario.
We note here that the RWPT model, being a 2D model cannot be applied to the membrane system.

These observations can also be understood by appropriate extension of our approximative model.
For this, we consider the membrane to consist of $n$ identical fibers each having a random orientation. 
Being close to each other, one can assume that all fibers share the same pool of potential reacting molecules.  
Thus, the efficiency of the $i^{th}$ fiber from the membrane can be regarded as the efficiency of a single fiber applied to the unreacted molecules left over from the previous $(i-1)$ fibers. 
Adding up all the $n$ fibers, the efficiency of the membrane can thus be written as:
\begin{equation}
\epsilon_\mathrm{mem}=\sum_{i=1}^{n}\epsilon\left(1-\epsilon\right)^{i-1} 
\label{eff_memb1}
\end{equation}
Computing the sum, we find
\begin{equation}
\epsilon_\mathrm{mem}=\epsilon\frac{\left(1-\epsilon\right)^{n}-1}{\left(1-\epsilon\right)-1}
\label{eff_memb2}
\end{equation}
which yields the final form for a $n$-fiber membrane
\begin{equation}
\epsilon_\mathrm{mem}=1-\left(1-\epsilon\right)^{n}
\label{eff_memb}
\end{equation}
As can be seen also in Fig.~\ref{figABmemb}, this model extension is in similarly good agreement with the simulation data as was already the case the isolated fiber.

%
% Cascade
%

\subsection{Cascade reaction}
\label{sec:res_cascade}
In order to investigate the $A\rightarrow B\rightarrow C$ cascade reaction, a second fiber (or membrane) responsible for the conversion of $B$ into $C$ is introduced into our simulation and theoretical model. 
The additional parameter compared to the single reaction is the distance $ \xi $ between the two catalysts which, in the limit $ \xi \to 0$, yields the side-by-side morphology illustrated in Fig.~\ref{system_ABC}.

\subsubsection{Pure diffusion case}
\label{sec:res_diff}
\begin{figure} [h]
	%single column image
	\includegraphics[width=0.8\linewidth]{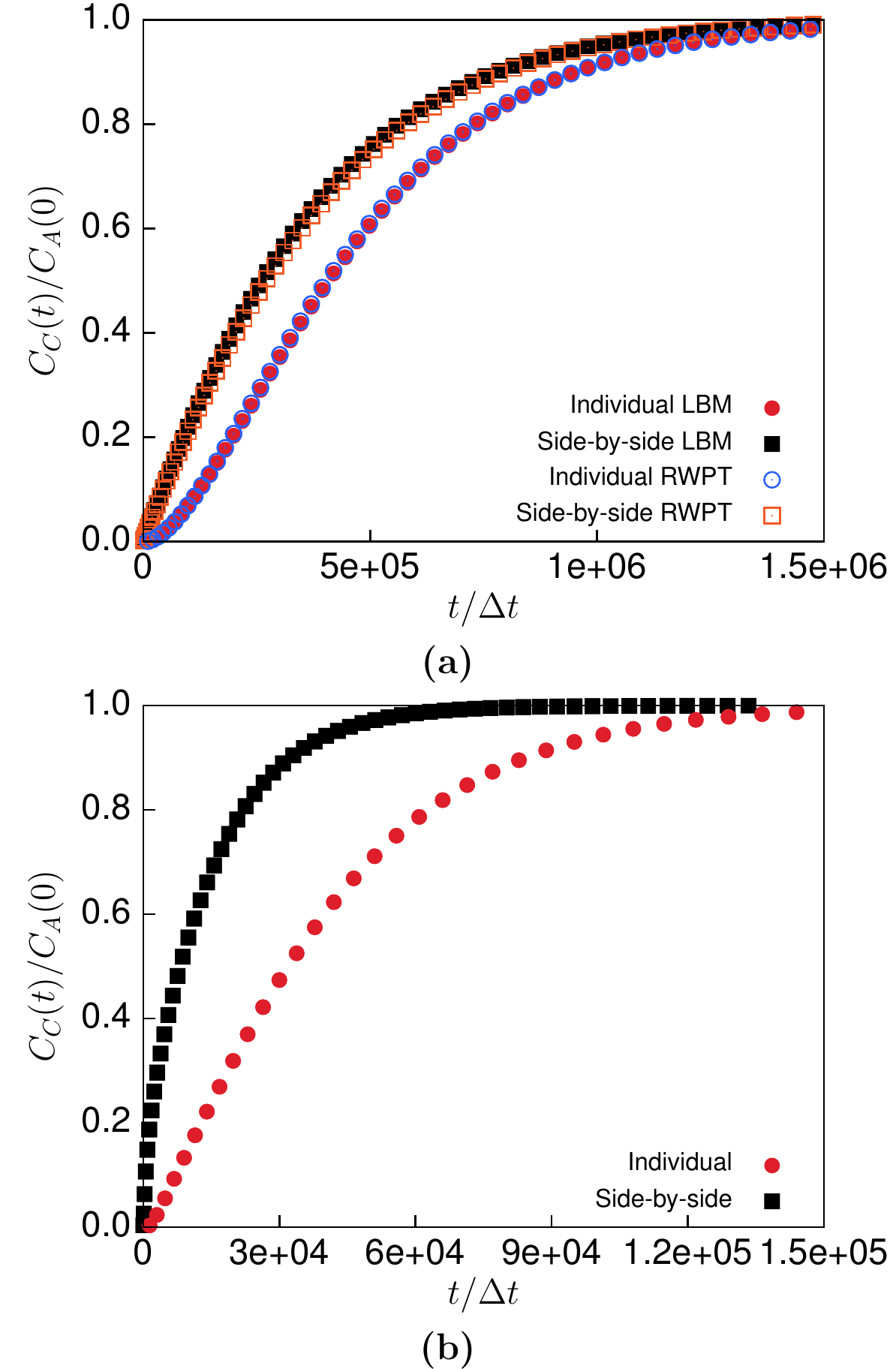}
	\centering
	\caption{Reaction progress in the absence of external flow. 
		(a) The comparison between two fibers separated by $\xi=100$ and the side-by-side morphology (both illustrated in Fig.~\ref{system_ABC}(a) and~(b), respectively) shows a slight superiority of the latter. 
		(b) A similar effect is observed for the membrane geometry illustrated in Fig.~\ref{system_ABC}(c) and~(d).
		All curves for $\mathrm{Da}=150$, corresponding data at $\mathrm{Da}=1.5$ is shown in Fig.~\ref{noflow_k001}}
	\label{noflow}
\end{figure}

To assess more clearly the influence of the catalyst distance, we start by investigating a slightly modified simulation setup: at the start of the simulation all space is filled homogeneously with species $A$, external flow is absent and periodic boundaries in all directions are imposed.
Instead of the efficiency $ \epsilon$ in the steady state, we monitor the total concentration of the three species over time.
To study the single fibers, a $300\times300$ box was used and two fibers with a radius $R=5$ were placed far away from each other for the individual approach, while for the side-by-side morphology a single fiber with a radius $R=10$ was considered. 
For the randomly generated membranes, the box size was set to $200\times60\times60$ and six fibers were used to form a membrane having $R=3$ and $R=6$ for the individual and side-by-side, respectively. 
Fig.~\ref{noflow}a shows the corresponding data for two fibers separated by a distance $\xi=100$ compared to the side-by-side morphology.
Even though the total surface areas for both systems were kept constant, the side-by-side morphology proved to be faster due to the placement of the catalysts next to each other such that the intermediate species required less time to reach the next catalytic site.

% Cascade: advection-diffusion
\subsubsection{Advection-Diffusion case}
\label{sec:res_advdiff}
\begin{figure} [h]
	%single column image
	\includegraphics[width=0.8\linewidth]{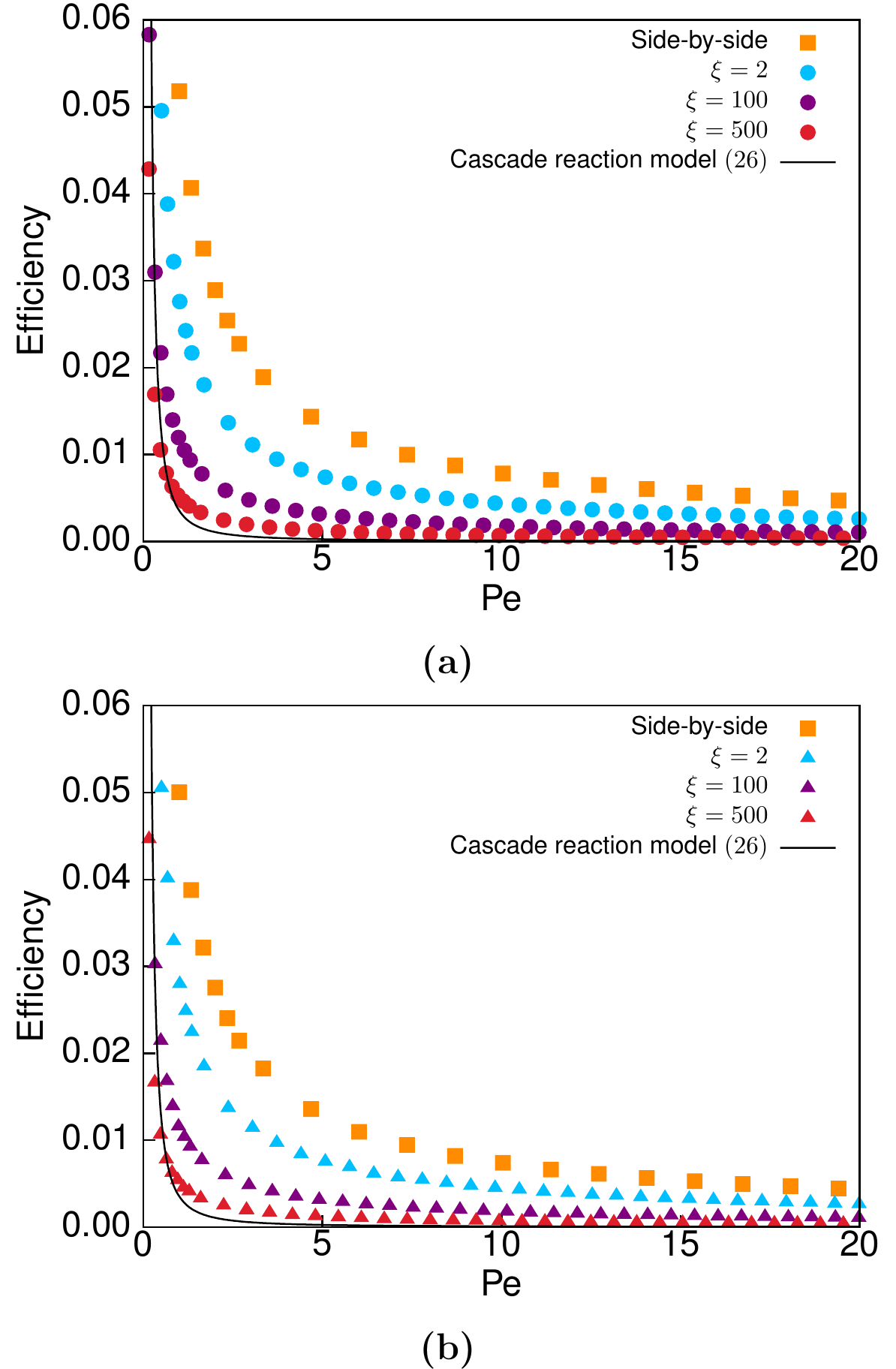}
	\centering
	\caption{Efficiencies of $A\rightarrow B\rightarrow C$ cascade reactions with $\mathrm{Da}=150$ using (a) LBM and (b) RWPT method. 
		The system geometry with fibers at different distances $\xi$ is illustrated in Fig.~\ref{system_ABC}(a).
		Simulation box is $1000\times300$ with cylinder radius $R=5$ for the individual fibers and $R=10$ for the side-by-side}
	\label{ABCdist}
\end{figure}

We now return to the flow-through reactor setup.
Fig.~\ref{ABCdist}(a) and~(b) show LBM and RWPT simulation data for the efficiency $\epsilon$ as function of $\mathrm{Pe}$ for two catalytic fibers separated by a distance $\xi$.
We observe the same trend as in the previous section, namely that a closer spacing leads to more efficient reactions with the highest $\epsilon$ achieved for the side-by-side morphology.
For completeness, we note that the relatively large difference between the $\xi=2$ and the side-by-side scenario is in part due to the definition of $\mathrm{Pe}$ which involves the differing radii $R=5$ and $R=10$.
Nevertheless, even when plotted as a function of the absolute flow velocity, the side-by-side scenario remains the most efficient geometry (see Fig.~\ref{ABCdist_velocity}).

We now aim to extend our theoretical model to include the second catalyst. 
For this, we start by placing the second fiber downstream at a distance $\xi>2d$ from the first one such that the two collision zones do not overlap.
The expected efficiency is then given by the product of the efficiency for each individual fiber
\begin{equation}
\epsilon_\mathrm{ABC}=\epsilon^{2}.
\label{eps_ABC}
\end{equation}
As shown by the comparison in Fig.~\ref{ABCdist}, this approach indeed reproduces nicely the simulation data at large $ \xi $.

\begin{figure}
	%single column image
	\includegraphics[width=0.8\linewidth]{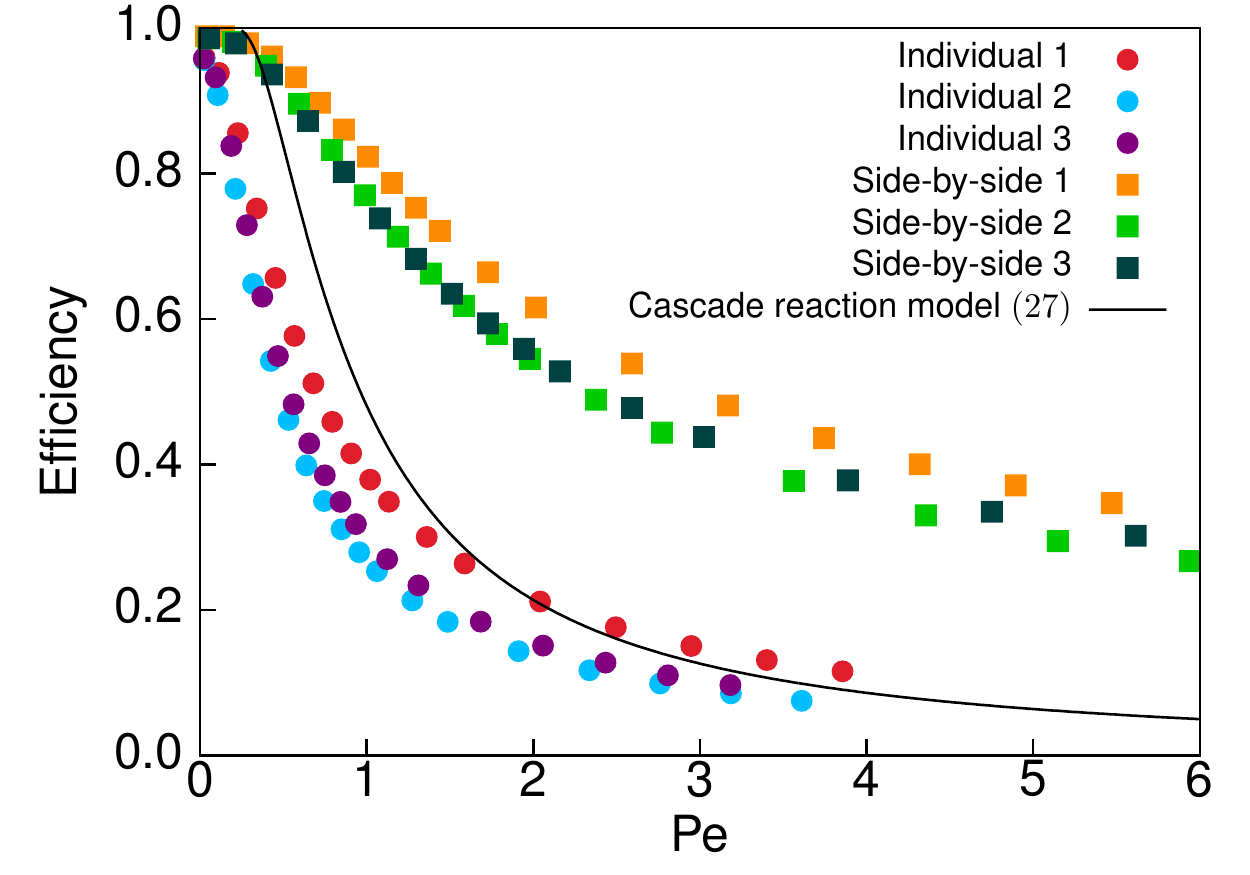}
	\centering
	\caption{Efficiencies of randomly generated membranes for the $A\rightarrow B\rightarrow C$ reaction. 
		All three sets of membranes for each morphology show good agreement with the proposed models, despite the various arrangements}
	\label{ABC_memb}
\end{figure}

In order to simulate the cascade reaction for a multi-fiber system, two membranes carrying each catalyst were assembled using six fibers with a radius $R=3$, as shown in Fig.~\ref{system_ABC}(c). 
For comparison, the side-by-side membrane consisting of six fibers but with a radius $R=6$ was also studied (Fig.~\ref{system_ABC}(d)). 
Three different random configurations were used. 
In complete analogy to equation~(\ref{eps_ABC}), our model predicts the total efficiency in the form
\begin{equation}
\epsilon_\mathrm{mem, ABC} = \epsilon_\mathrm{mem}^2
\label{eff_ABC_memb}
\end{equation}
which is in good agreement with the simulation data as shown in Fig.~\ref{ABC_memb}. 
The agreement between the model and all three random membranes illustrates that the precise arrangement of fibers within a membrane is only of secondary importance for its catalytic efficiency.

\section{Conclusions}
\label{sec:conclusion}
We investigated the catalytic efficiency of fibrous membranes using Lattice-Boltzmann simulations of the advection-diffusion-reaction equations as well as an approximative analytical model.
Starting with one-step $A\to B$ reactions, our main focus then was on cascade $A\to B \to C$ reactions where two fiber systems with different catalysts are required.
The control parameters of the system, besides its geometry, can be encapsulated into only two non-dimensional numbers: (i) the P\'eclet number relating advection and diffusion and (ii) the Damk\"ohler number relating reaction and diffusion.
Our simulations allowed us to compute the system efficiency of a flow-through reactor for a large set of parameters and geometries.
Our theoretical model, containing only a single adjustable parameter $\Delta$ turned out to be in full agreement with the numerical simulations.

\section*{Acknowledgements}
This project was funded by the Deutsche Forschungsgemeinschaft, SFB 840 (subproject A12) and the Volkswagen Foundation.
We gratefully acknowledge computing time provided by the SuperMUC system of the Leibniz Rechenzentrum and the Bavarian Polymer Institute.

\appendix
\section{Production rate}
\begin{figure*} [h]
	%2-column image
	\includegraphics[width=1\textwidth]{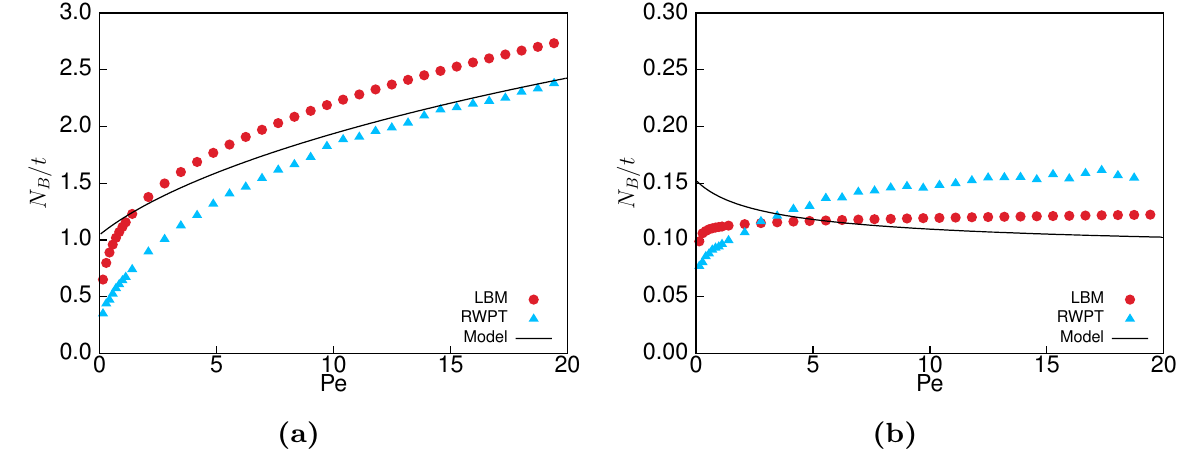}	%R=5
	\centering	
	\caption{Production rate of $B$ species as a function of $\mathrm{Pe}$ for Fig.~\ref{singleAB}.
		Parameters are (a) $\mathrm{Da=150}$ ($k=1$) and (b) $\mathrm{Da=1.5}$ ($k=0.01$)}
	\label{prod_AB_single}
\end{figure*}

\begin{figure*} [h]
	%2-column image
	\includegraphics[width=1\textwidth]{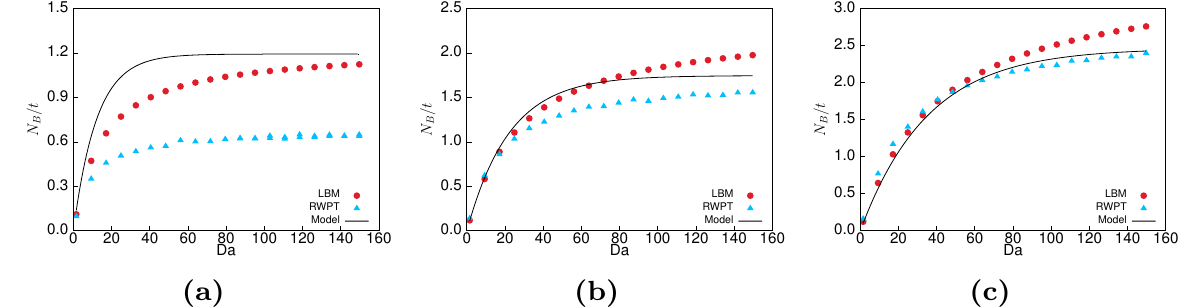}	%R=5
	\centering	
	\caption{Production rate of an $A\rightarrow B$ reaction with an isolated fiber as a function of Da for (a) $\mathrm{Pe}=1$, (b) $\mathrm{Pe}=7$, (c) $\mathrm{Pe}=20$ corresponding to Fig.~\ref{singleAB_Da}}
	\label{prod_AB_single_Da}
\end{figure*}

\begin{figure*} [h]
	%2-column image
	\includegraphics[width=1\textwidth]{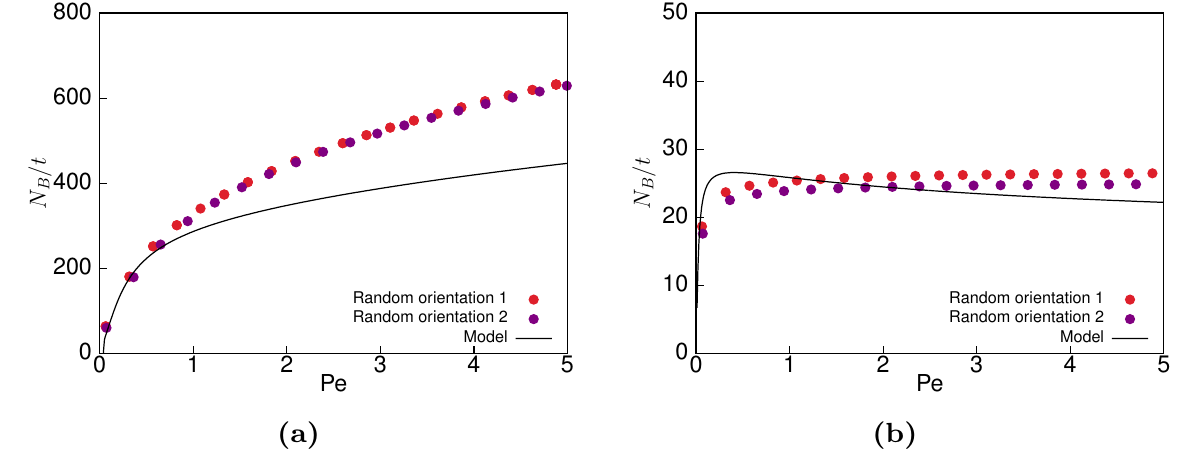}
	\centering
	\caption{Production rate for Fig.~\ref{figABmemb}. (a) $\mathrm{Da}=150$, (b) $\mathrm{Da}=1.5$}
	\label{prod_AB_memb}
\end{figure*}

\begin{figure*} [ht]
	%2-column image
	\includegraphics[width=1\textwidth]{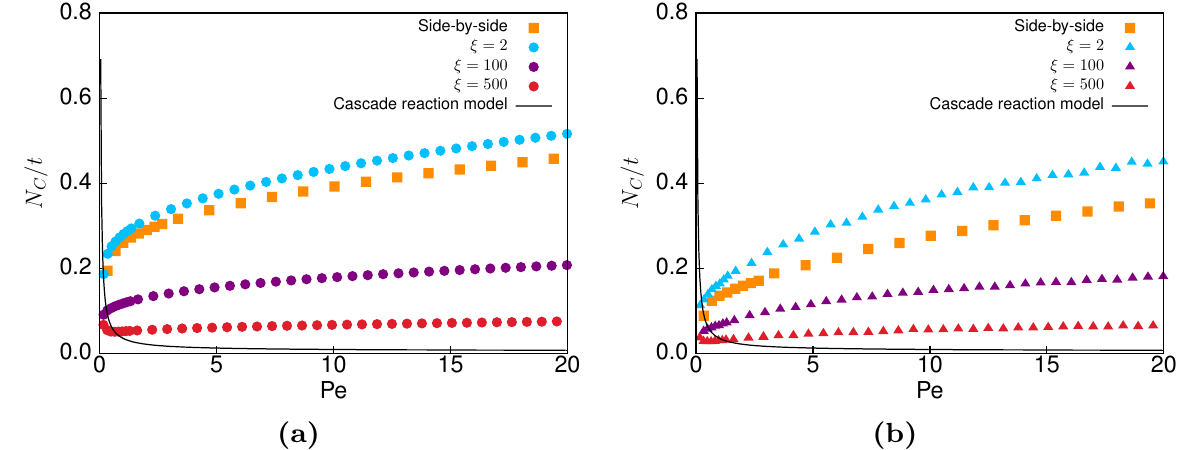}
	\centering
	\caption{Production rate for Fig.~\ref{ABCdist} (cascade reaction with $\mathrm{Da}=150$) using (a) LBM and (b) RWPT method. 
		Note that the different order of the orange/blue curves compared to the main text is again due to Pe involving fiber radii}
	\label{prod_ABCdist}
\end{figure*}

\begin{figure*} [ht]
	%2-column image
	\includegraphics[width=0.5\linewidth]{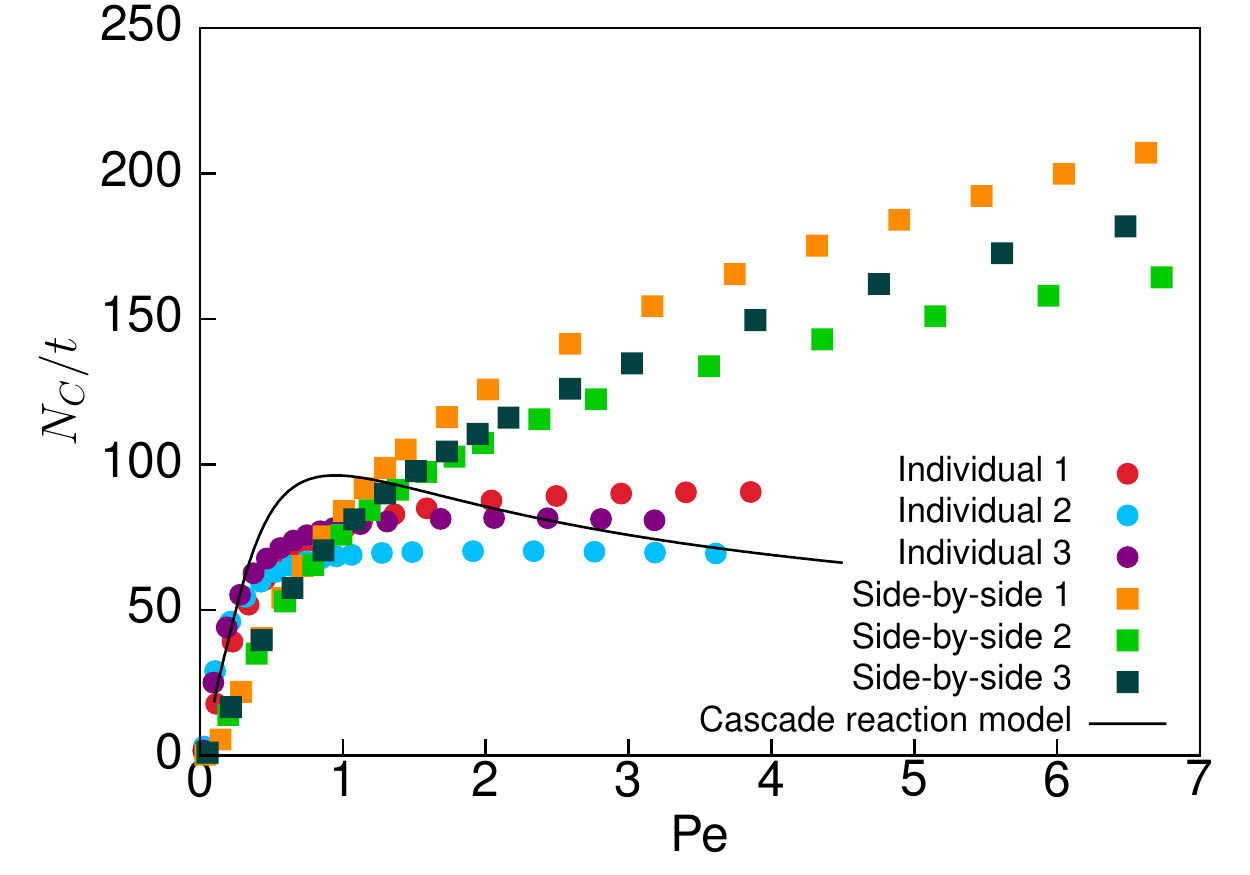}
	\centering
	\caption{Production rate for Fig.~\ref{ABC_memb}. All curves for $\mathrm{Da}=150$}
	\label{prod_ABC_memb}
\end{figure*}

Besides the efficiency, another quantity that may be of interest when analyzing a flow-through reactor is the production rate. 
For this, we consider the amount of final product that leaves the reactor per unit time. 
As Figs.~\ref{prod_AB_single}-\ref{prod_ABC_memb} show, despite the decrease in efficiency at high $\mathrm{Pe}$, the higher throughput leads to an increase of the production rate of the final species for almost all cases.

%%
%% Additional figures
%%
\clearpage

\section{Additional data}
\begin{figure} [htb]
	%single column image
	\includegraphics[width=0.8\linewidth]{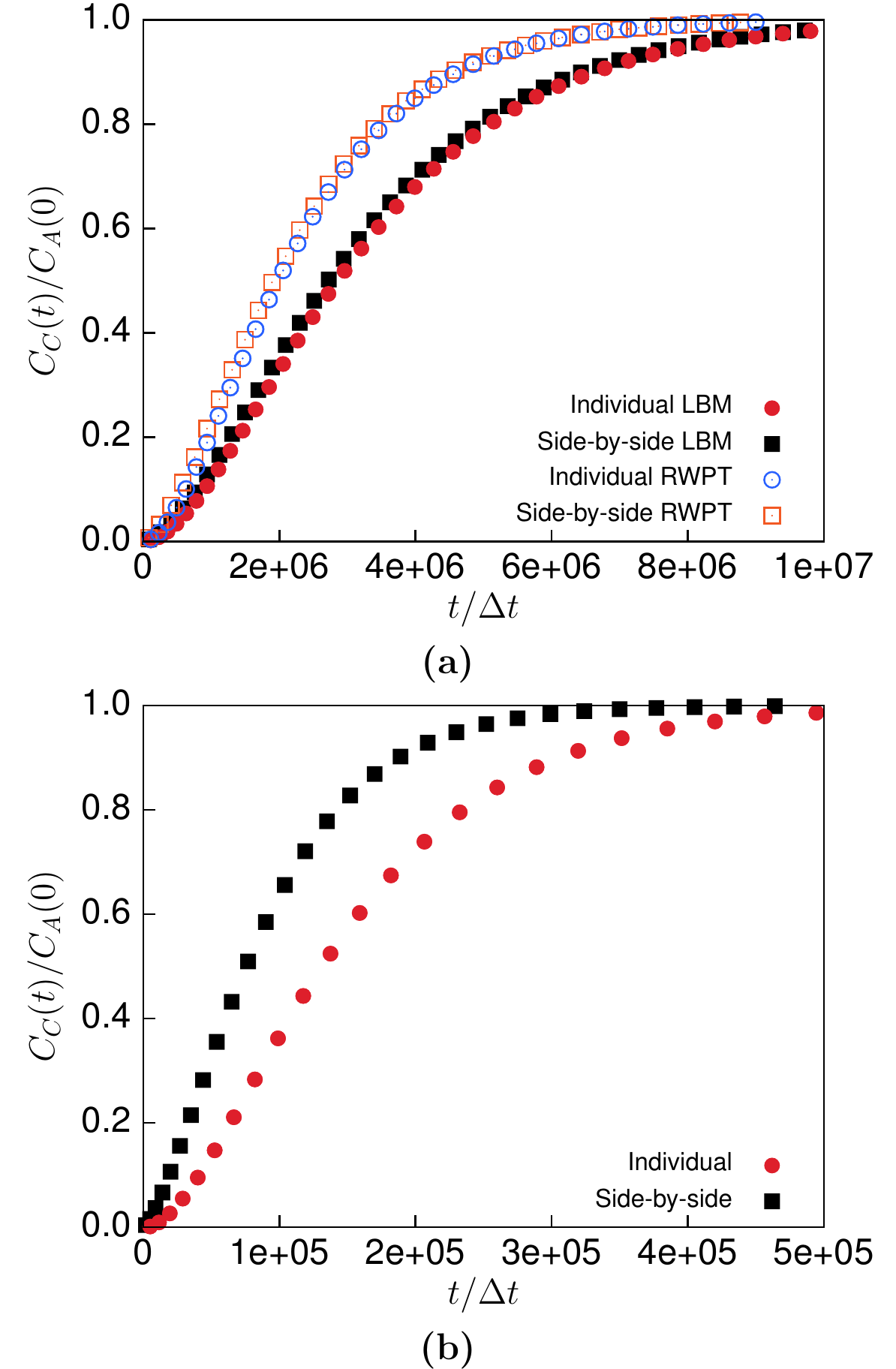}
	\centering
	\caption{As Fig.~\ref{noflow}, but with $\mathrm{Da}=1.5$}
	\label{noflow_k001}
\end{figure}

\begin{figure} [h]
	%single column image
	\includegraphics[width=0.8\linewidth]{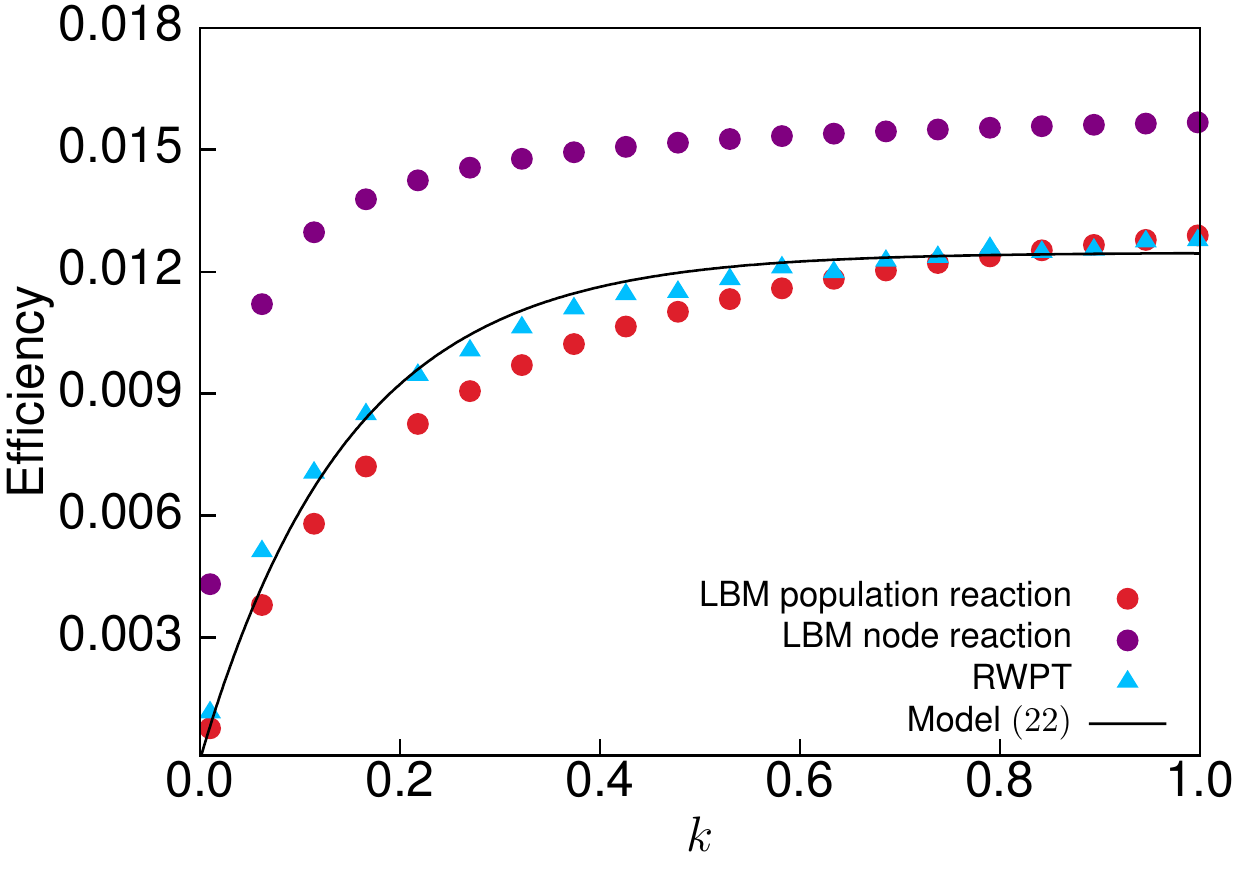}
	\centering	
	\caption{As Fig.~\ref{singleAB_Da}(b) but including an LBM simulation where all populations are included into the reaction term in equation~(\ref{source_lbm}).
		The agreement between LBM, RWPT and the theoretical demonstrates the correctness of the used approach}
	\label{node_vs_pop}
\end{figure}

\begin{figure} [h]
	%single column image
	\includegraphics[width=0.8\linewidth]{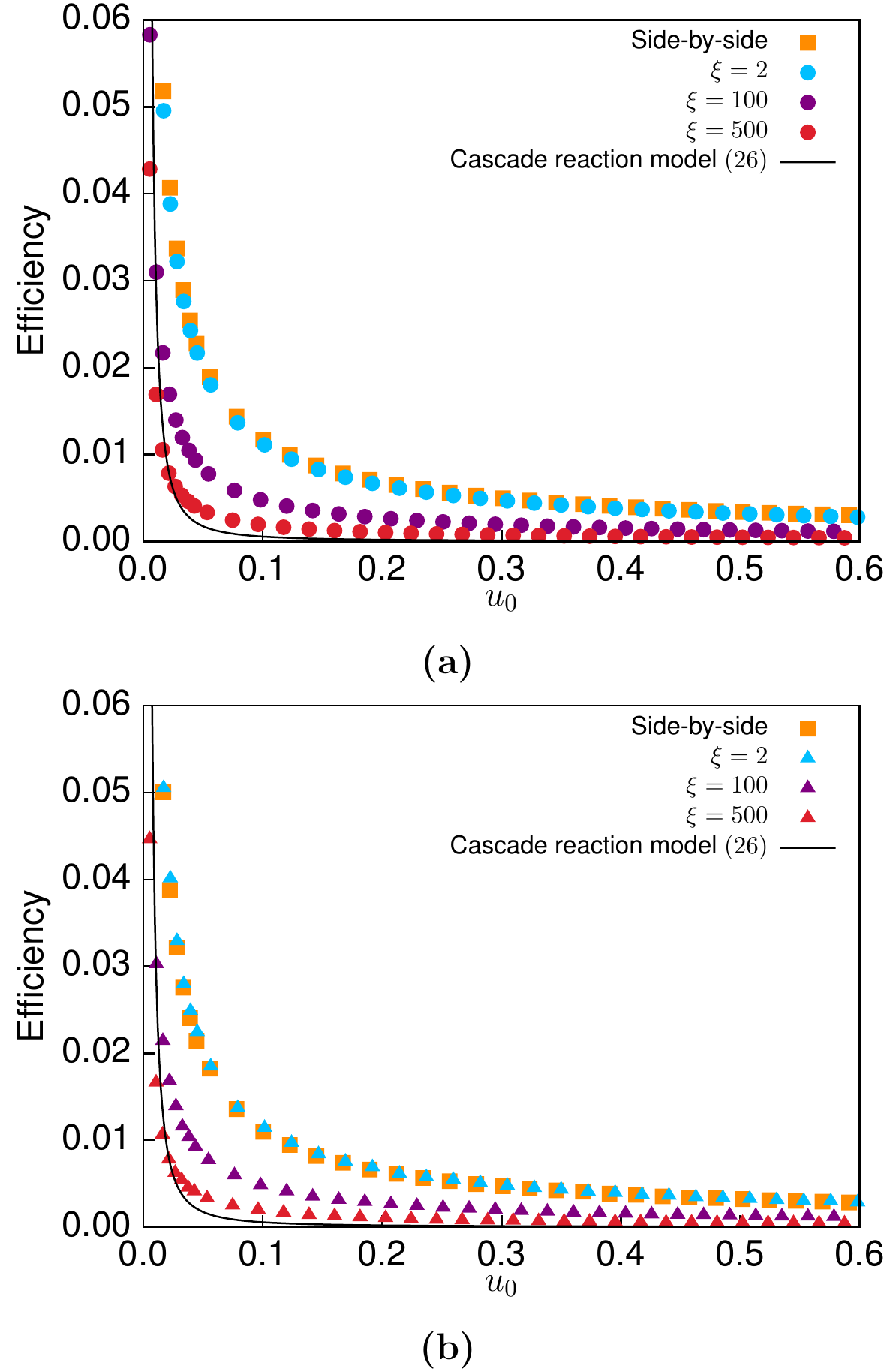}
	\centering
	\caption{ Same efficiencies as in Fig.~\ref{ABCdist}(a) and~(b), respectively, but plotted as a function of the absolute flow velocity}
	\label{ABCdist_velocity}
\end{figure}
\clearpage

%
% Validation
%

\section{Validation of LBM for ADRE}

%
% Gaussian hill
%

\subsection{Advection-Diffusion of a Gaussian Hill}

We consider the diffusion and advection of a species in a uniform velocity field $\vec{u}$ in a 2D system ($100\times1\times100$). 
Starting with a Gaussian concentration profile with the width $\sigma_{0}$ given by
\begin{equation}
C\left( \vec{x},t=0\right)=C_{0}\exp\left(-\frac{\left( \vec{x}-\vec{x_{0}}\right)^{2} }{2\sigma_{0}^{2}} \right)
\end{equation}
the results can be compared with the analytical solution \citep{antiBB}
\begin{equation}
C\left(\vec{x},t\right)=\frac{\sigma_{0}^{2}}{\sigma_{0}^{2} + \sigma_{D}^{2}}C_{0} \exp\left(-\frac{\left( \vec{x} - \vec{x_{0}} - \vec{u} t\right)^{2} }{2 \left( \sigma_{0}^{2} + \sigma_{D}^{2} \right)  } \right) 
\end{equation}
where $\sigma_{D}^{2}=2Dt$. 
We set the initial concentration $C_{0}=1$ and place the Gaussian hill with the width $\sigma_{0}=\Delta x$ in the center of the domain. 
As in all our simulations, we use here $\Delta x=1$, $\Delta t=1$ and $\tau_ \mathrm{g}=\Delta t$ in equation~(\ref{Diffusion_LBM}) resulting in a diffusion coefficient $D=0.1666$.

\begin{figure} [h]
	%single column image
	\includegraphics[width=0.8\linewidth]{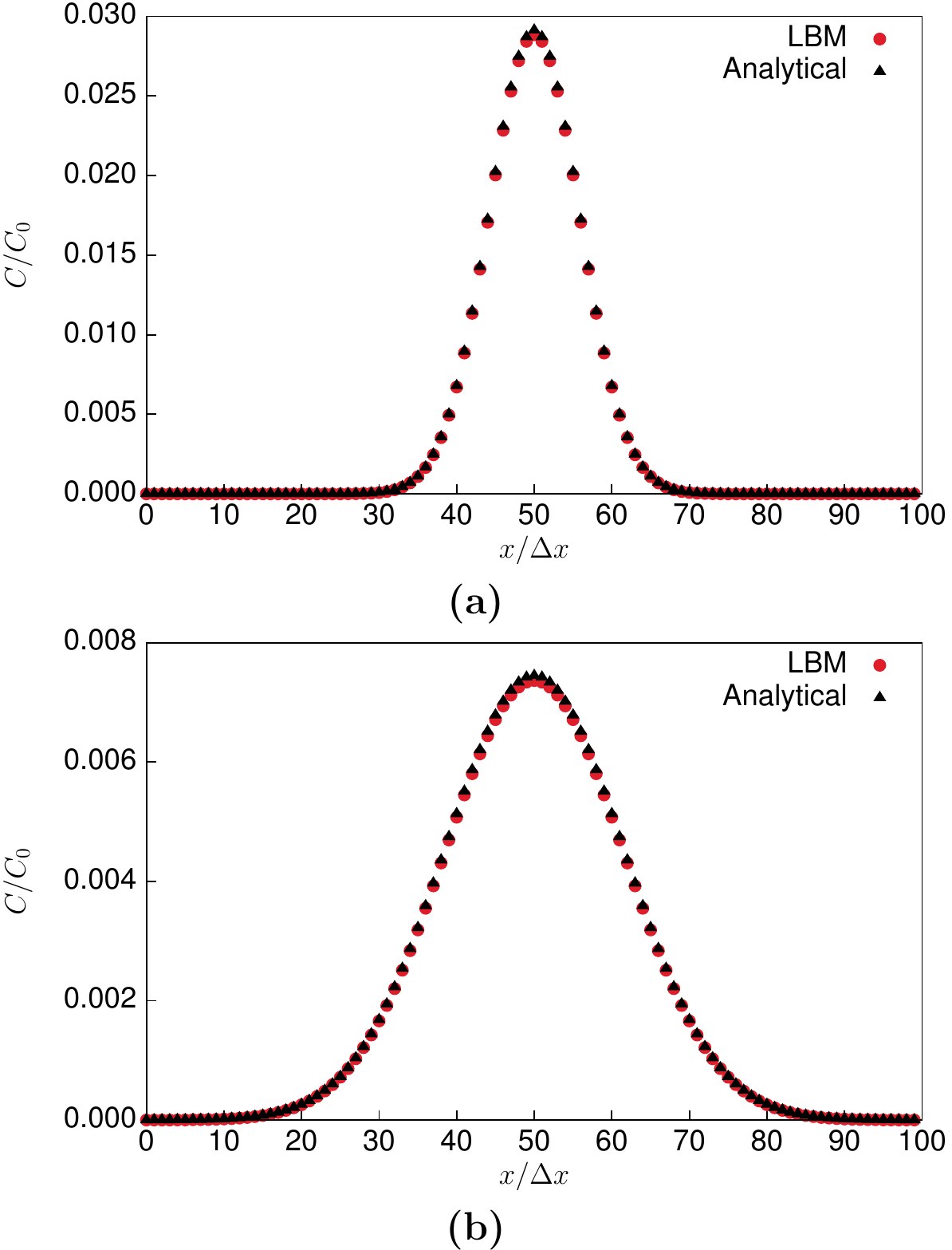}
	\centering
	\caption{Concentration profile of the Gaussian hill in the pure diffusion regime at (a) $t=100\Delta t$ and (b) $t=400\Delta t$}
	\label{profile_diff}
\end{figure}

\begin{figure} [h]
	%single column image
	\includegraphics[width=0.8\linewidth]{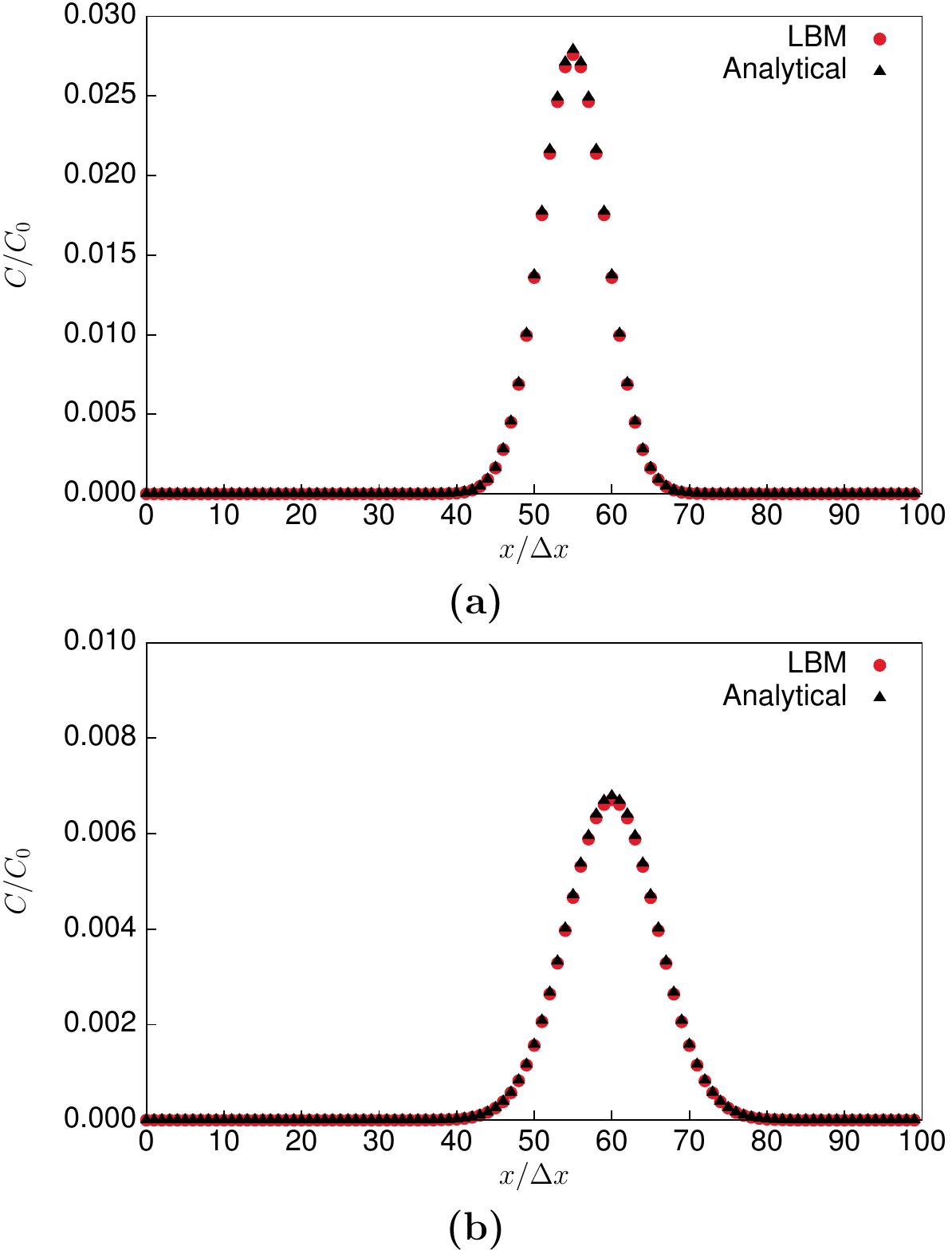}
	\centering		
	\caption{Concentration profile of the Gaussian hill in the advection-diffusion regime at (a) $t=50\Delta t$ and (b) $t=100\Delta t$}
	\label{profile_adv}
\end{figure}

First, we consider only the diffusive regime ($\mathrm{Pe}=0$) by choosing $\vec{u}=\vec{0}$. Fig.~\ref{profile_diff} illustrates that our model matches the analytical results and the small difference between the two does not increase over time.
By introducing a velocity $\vec{u}=(0.1,0.1) \Delta x/\Delta t$, the previous profile starts to shift (Fig.~\ref{profile_adv}). Again, a very good agreement between our simulations and the theoretical results is obtained.

%
% Latini
%

\subsection{Advection-diffusion in a microchannel}

A more challenging benchmark problem than the Gaussian hill is to replicate the three regimes of dispersion for a point discharge of tracer particles in laminar pipe flow as calculated in \cite{latini_2001}.
Starting from a $\delta$-function initial concentration at the center of the pipe, the moments of the longitudinal distribution of the tracer, $M_n(t)$, can be computed according to equation (1.5) of \citep{latini_2001}. 
The width of the distribution is then defined as
\begin{equation}
\sigma(t)=\sqrt{M_2(t)-M_1^2(t)}.
\end{equation}
Analyzing its time dependency reveals the three stages of longitudinal dispersion. 
For small times, diffusion dominates ($\sigma \sim \sqrt{2D_\mathrm{Latini}t}$), followed by the anomalous regime where the width scales superdiffusively ($\sigma \sim \sqrt{8/3}D_\mathrm{Latini}t^2$). 
Finally, at large times, the flow enters the Taylor regime where the width scales diffusively again ($\sigma \sim \sqrt{2D_\mathrm{Latini}^\mathrm{eff}t}$), but with a larger diffusion coefficient $D_\mathrm{Latini}^\mathrm{eff}=\frac{1}{192D_\mathrm{Latini}+D_\mathrm{Latini}}$.

\begin{figure} [h]
	%single column image
	\includegraphics[width=0.8\linewidth]{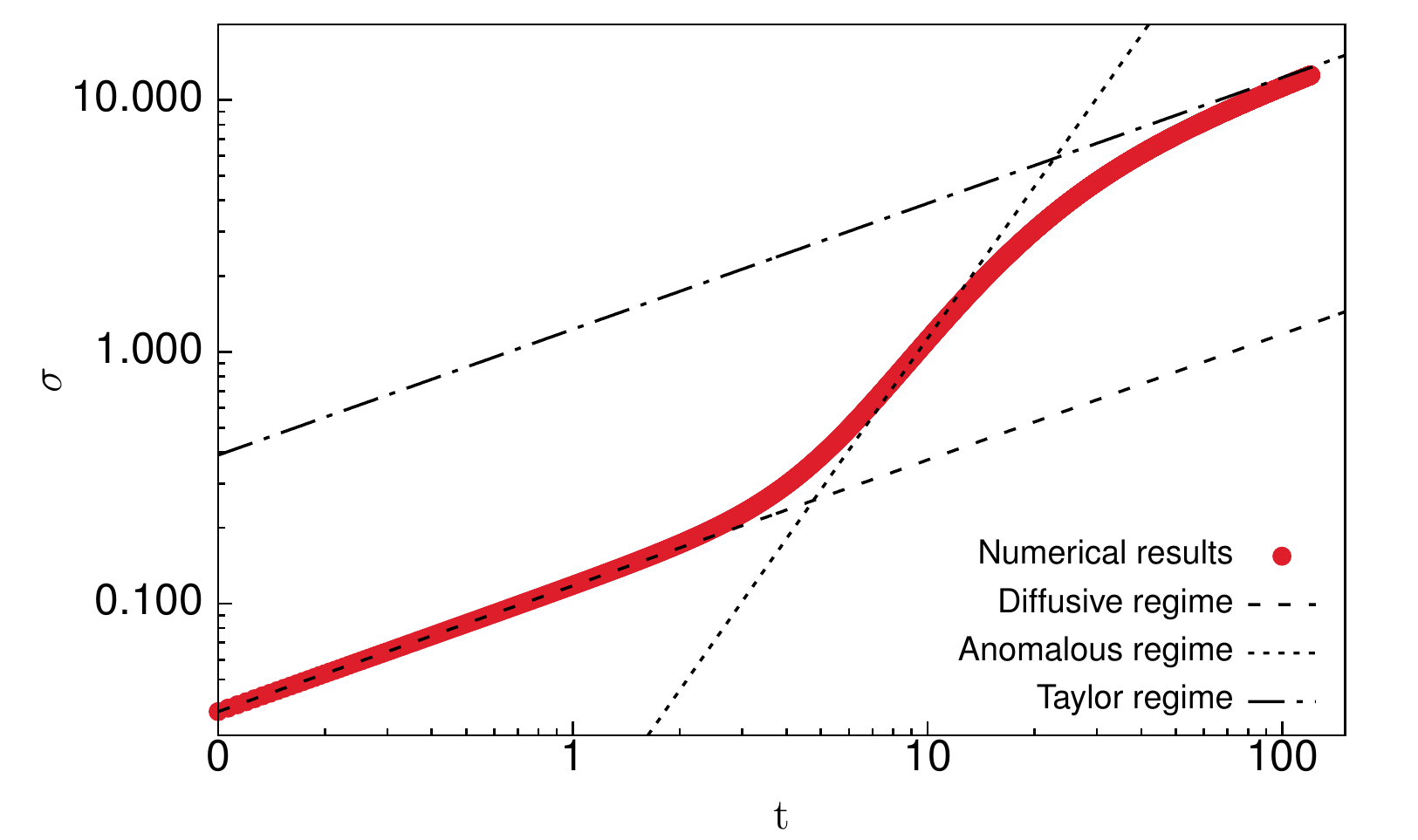}
	\centering	
	\caption{The three stages of longitudinal dispersion for $D_{Latini}=6.94\times10^{-3}$}
	\label{latini}
\end{figure}

A $20000 \times 123 \times 123$ grid was used to simulate a pipe with a radius $R=60$. 
The centerline velocity of the flow was set to $U_{0}=0.4$ and a relaxation time $\tau=\Delta t$ was used as above, thus fixing the dimensionless diffusion coefficient $D_\mathrm{Latini}=\frac{D}{RU_0}=6.94\times10^{-3}$. 
Computing the width of the distribution $\sigma$ as given in \cite{latini_2001} shows a very good agreement with the theoretical values especially for the diffusive and Taylor regimes (Fig.~\ref{latini}). 
In order to obtain a clear anomalous regime, a very large lattice must be used which requires a lot of memory.
The alternative would be to set the relaxation time close to $\tau=0.5\Delta t$, but that can lead to negative populations and is avoided here.

%
% Homogeneous reaction
%

\subsection{Homogeneous reaction}

\begin{figure} [ht]
	%single column image
	\includegraphics[width=0.8\linewidth]{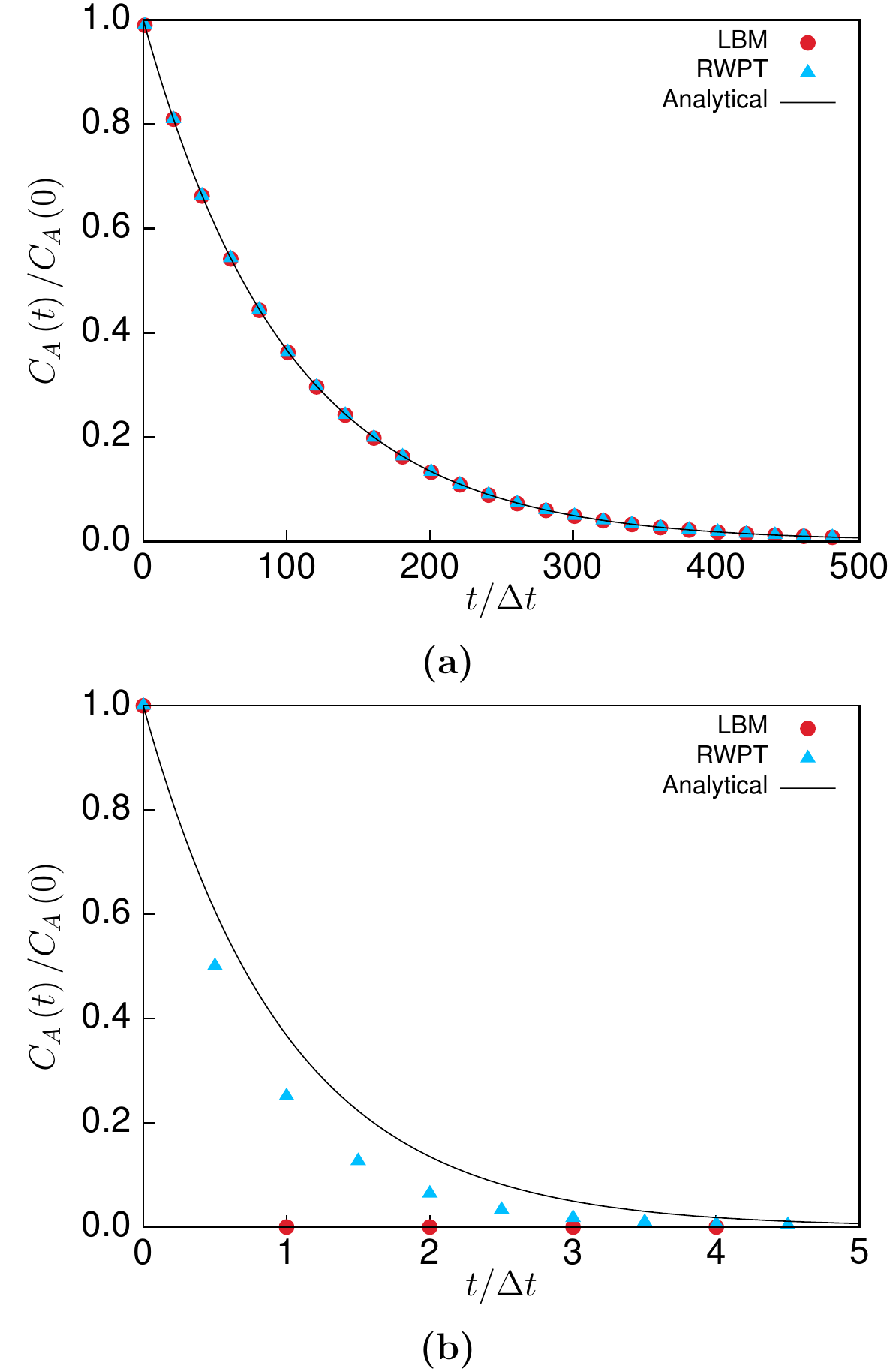}
	\centering
	\caption{Conversion curves for a first-order homogeneous reaction with the reaction rate (a) $k=0.01$ and (b) $k=1$
	}
	\label{homo_react}
\end{figure}

We consider a first-order $A \rightarrow B$ reaction in a periodic and homogeneous $50\times50\times50$ system. 
The well known rate law $C_A(t)=e^{-kt}C_A(0)$ is compared to LBM and RWPT simulations.
Fig.~\ref{homo_react} shows that our assumed form matches very well the theoretical model in the low $k$ regime and that for higher values a good compromise between speed and accuracy can be achieved by setting the time step $\Delta t=0.5$ for the particle based model.
The LBM model with $\Delta t=1$ decays to zero instantaneously as expected.

\clearpage

%\section*{References}
\bibliography{bibl}

\end{document}